%% file: main.tex
\documentclass[letterpaper,twocolumn,10pt]{article}
\usepackage{usenix2019_v3}

\usepackage[final,inline,nomargin,index]{fixme}
\usepackage[firstpage]{draftwatermark}
\fxsetup{theme=color,mode=multiuser}
\usepackage[ruled,vlined,linesnumbered]{algorithm2e}
\usepackage{algpseudocode}
\usepackage{amssymb}
\usepackage{amsthm}
\usepackage{amsmath}
\usepackage{appendix}
\usepackage{booktabs}
\usepackage{enumitem}
\usepackage{inconsolata}
\usepackage{listings}
\usepackage{multirow}
\usepackage{shortcuts}
\usepackage{subcaption}
\usepackage{tabularx}
\usepackage{threeparttable}
\usepackage{tikz}
\usepackage{url}
\usepackage{xspace}

\theoremstyle{definition}
\newtheorem{defn}{Definition}

\theoremstyle{definition}

\begin{document}

\title{
  \sysname: A Neural Network Language Model-Guided \\
  JavaScript Engine Fuzzer
}

\SetWatermarkText{
  \hspace*{6in}\raisebox{7.5in}{
    \includegraphics{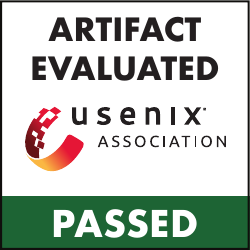}
  }
}
\SetWatermarkAngle{0}

\author{
{\rm Suyoung Lee, HyungSeok Han, Sang Kil Cha, Sooel Son}\\
School of Computing, KAIST \\
}

\maketitle
\pagestyle{empty}

\input{abstract}
\input{intro}
\input{background}
\input{motive}
\input{overview}
\input{design}
\input{implementation}
\input{experiment}
\input{related}

\input{conclusion}

\section*{Acknowledgements}
We thank anonymous reviewers for their helpful feedback
and Yale Song who helped develop some of the ideas used in
this paper.
We are also grateful to Jihoon Kim for kindly sharing his findings, which indeed
inspired our project.
Finally, we thank Sunnyeo Park for collecting JS seeds used for the evaluation.
This work was partly supported by (1) Institute for Information \&
communications Technology Promotion (IITP) grant funded by the Korea government
(MSIT), No.2018-0-00254, and (2) LIG Nex1.

\input{auth.bbl}
\end{document}

%% file: abstract.tex
\begin{abstract}
  JavaScript (JS) engine vulnerabilities pose significant security threats
  affecting billions of web browsers. While fuzzing is a prevalent technique for
  finding such vulnerabilities, there have been few studies that leverage the
  recent advances in neural network language models (NNLMs). In this paper, we
  present \sysname, the first NNLM-guided fuzzer for finding JS engine
  vulnerabilities.
  The key aspect of our technique is to transform a JS abstract syntax tree
  (AST) into a sequence of AST subtrees that can directly train prevailing
  NNLMs. We demonstrate that \sysname is capable of generating valid JS tests,
  and show that it outperforms previous studies in terms of finding
  vulnerabilities. \sysname found \ourbug real-world bugs, including three CVEs,
  in the latest JS engines, demonstrating its efficacy in finding JS engine bugs.

\end{abstract}

%% file: intro.tex
\section{Introduction} \label{s:intro}

The memory safety of web browsers has emerged as a critical attack vector as
they have become an integral part of everyday computing. Malicious websites,
which conduct drive-by download attacks~\cite{adityad:bd-2016}, have typically
exploited memory corruption vulnerabilities of web browsers.
Currently, an exploitable memory corruption vulnerability for a browser can cost
100,000 USD and sell for a million dollars if it is chained with a kernel
exploit to remotely jailbreak iOS~\cite{zerodium}.

Among many components of web browsers, a JavaScript (JS) engine is of
particular interest to attackers as its Turing-complete nature enables attackers
to craft sophisticated exploits. One can easily allocate a series of heap chunks
to perform heap spraying~\cite{sotirov:blackhat:2007}, write functions in JS
to abstract away some exploitation logic~\cite{pwnjs}, and even bypass
the mitigation used in modern web browsers~\cite{molinyawe:blackhat:2016}.
According to the National Vulnerability Database (NVD), \percentCVE of the total
vulnerabilities reported for Microsoft Edge and Google Chrome in 2017
were JS engine vulnerabilities.

Despite the increasing attention, there has been relatively little academic
research on analyzing JS engine vulnerabilities compared to other
studies seeking to find them~\cite{holler:usec:2012, guo:2013,
veggalam:esorics:2016}.
LangFuzz~\cite{holler:usec:2012} combines code fragments extracted from
JS seed files to generate JS test inputs. GramFuzz and IFuzzer employ more or
less the same approach~\cite{guo:2013, veggalam:esorics:2016}, but
IFuzzer uses evolutionary guidance to improve the fuzzing effectiveness with
genetic programming based on the feedback obtained by executing a target JS
engine with produced inputs.

However, none of the existing approaches consider the relationship between code
fragments for generating test inputs. In other words, they produce test inputs
by simply combining fragments as long as JS grammars allow it. Thus, they do not
determine which combination is likely to reveal vulnerabilities from the target JS
engine. Are there any similar patterns between JS test inputs that trigger
JS engine vulnerabilities? If so, can we leverage such patterns to drive
fuzzers to find security vulnerabilities? These are the key questions that
motivated our research.

We performed a preliminary study on JS engine vulnerabilities and observed
two patterns.
We observed that \emph{a new security problem often arises from JS engine
files that have been patched for a different bug.}
We analyzed 50 CVEs assigned to \chakra, a JS engine used by Microsoft Edge.
We found that \chakraGlobalOptPatchRate and \chakraJSArrayPatchRate of the
vulnerabilities were related to \texttt{GlobOpt.cpp} and \texttt{JavascriptArray.cpp},
respectively.

The second observation was that \emph{JS test code that triggers new security
vulnerabilities is often composed of code fragments that already exist in
regression tests.}
We collected \chakraNumRegressionTests unique JS files from the \chakra
regression test suite and \chakraNumPoC JS files that invoked the analyzed
vulnerabilities. These two sets of files were disjoint.
We sliced the AST of each JS file into AST subtrees of depth one, called
\textit{fragments}. We then computed the number of overlapping fragments between
the two sets; we found that \chakraAvgOverlapRate of the fragments extracted
from the \chakraNumPoC vulnerability-triggering JS files overlapped with the
fragments extracted from the regression test suite (see \S\ref{s:motive}).

Given these two observations, how do we perform fuzz testing to find JS engine
vulnerabilities?
For this research question, we propose the first approach that leverages
a \emph{neural network language model} (NNLM) to conduct fuzz testing on
a target JS engine.
Our key idea is to mutate a given regression JS test by replacing its partial
code with new code that the NNLM creates.
Consider a regression JS test that invokes a patched functionality. We generate
a JS test from this regression test while expecting to elicit a new potential
bug that resides in the patched JS engine files, thus addressing the first
observation.
We also assemble existing code from regression test suites under the guidance
of the NNLM when composing new partial code. This captures the second
observation.

To manifest this idea, we designed and implemented \sysname, a system for
finding security vulnerabilities in JS engines.
The system starts by transforming the AST of each JS test from a given
regression test suite into the sequence of fragments. These fragment sequences
become training instances over which the NNLM is trained. Therefore, the NNLM
learns the relationships between fragments. \sysname mutates a given JS test
by reconstructing one of its subtrees as the trained NNLM guides.

Previous research focused on learning the relationships between
PDF objects~\cite{godefroid:ase:2017}, characters~\cite{cummins:issta:2018,
liu:aaai:2019}, and lexical tokens in the source code~\cite{nguyen:fse:2013,
hindle:icse:2012, raychev:pldi:2014}.
These language models addressed completing incorrect or missing tokens~\cite{
tu:fse:2014, nguyen:fse:2013}, or assembling PDF objects~\cite{godefroid:ase:2017}.
Their methods are not directly applicable to generating valid JS tests, which
requires modeling structural control flows and semantic data dependencies among
JS lexical tokens.
Liu~\etal~\cite{liu:aaai:2019} stated their limitation in extracting general
patterns from character-level training instances from C code, thus generating
spurious tests.

Unlike these previous studies~\cite{cummins:issta:2018, godefroid:ase:2017},
\sysname uses fragments as building blocks. Each fragment encapsulates the
structural relationships among nodes within an AST unit tree.
The model is then trained to learn the relationships between such AST unit
trees. \sysname uses this model to assemble unit subtrees when mutating a given
regression JS test. Thus, each generated JS test reflects the syntactic and
semantic commonalities that exist in the regression test suite.

We evaluated \sysname to find bugs in \chakra 1.4.1 and compared
the number of found bugs against \alchemist~\cite{han:ndss:2019},
\jsfunfuzz~\cite{funfuzz}, and \ifuzzer~\cite{veggalam:esorics:2016}. We
performed five fuzzing campaigns; each round ran for 72 hours. \sysname found
133 bugs, including 15 security bugs. Among the found security bugs, \sysname
reported 9, 12, and 12 bugs that \alchemist, \jsfunfuzz, and \ifuzzer did not find, respectively.
This result demonstrates that \sysname is able to find bugs that the
state-of-the-art JS fuzzers are unable to find.

We measured the efficacy of the \sysname language model against the random
selection method with no language model, Markov-chain model, and the
character/token-level recurrent neural network language model. \sysname
outperformed the other approaches in terms of finding unique bugs.

We further tested \sysname to fuzz the latest versions of \chakra, \jsc, \moz,
and \veight. \sysname found \ourbug unique bugs, including three security bugs.
34 bugs were found from \chakra. The remaining two and one bugs were from \jsc
and \veight, respectively. Of these three security bugs, \sysname discovered
one from \jsc and the other two from \chakra.
These results demonstrate the effectiveness of leveraging NNLMs in finding
real-world JS engine bugs.

%% file: background.tex
\section{Background}
\label{s:background}

\subsection{Language Model}

A language model is a probability distribution over sequences of words. It is
essential for natural language processing (NLP) tasks, such as speech
recognition, machine translation, and text generation.
Traditionally, language models estimate the likelihood of a word sequence given
its occurrence history in a training set.

An $n$-gram language model~\cite{chen:amacl:1996, kim:aaai:2016}
approximates this probability based on the occurrence history of the
preceding $n-1$ words.
Unfortunately, such count-based language models inherently suffer from the
\emph{data sparsity problem}~\cite{chen:amacl:1996}, which causes them to
yield poor predictions. The problem is mainly due to insufficient representative
training instances.
NNLMs address the data sparsity problem by representing words as a distributed
vector representation, which is often called a \emph{word embedding}, and using
it as input into a neural network.

Bengio~\etal~\cite{bengio:npl:2003} introduced the first NNLM, a feed-forward
neural network (FNN) model. An FNN predicts the next word based on its preceding
$n-1$ words, which is called a \textit{history} or a
\textit{context} where $n$ is a hyper parameter that represents the size of the
word sequence~\cite{ bengio:npl:2003,goodman:csl:2001,arisoy:wlm:2012}.
In this NNLM setting, all words in a training set constitute a vocabulary $V$.
Each word in $V$ is mapped onto a feature vector. Therefore, a \textit{context},
a word sequence, becomes the concatenation of each feature vector corresponding
to its word. The model is then trained to output a conditional probability
distribution of words in V for the next word from a given \textit{context}.

\noindent\textbf{Long short-term memory (LSTM).}
Unlike FNN language models, a recurrent neural network (RNN) is capable of
predicting the next word from a history of preceding words of an arbitrary length
because an RNN is capable of accumulating information over a long history of words.
An LSTM model is a special kind of RNN; it is designed to capture long-term
dependencies between words~\cite{hochreiter:nc:1997, gers:lstm:1999}. Because a
standard RNN suffers from the gradient vanishing/exploding problem~\cite{
bengio:tnn:1994}, an LSTM model uses neural layers called gates
to regulate information propagation and internal memory to update its
training parameters over multiple time steps.

\subsection{JS Engine Fuzzing} \label{ss:jsfuzzing}

Fuzz testing is a form of dynamic software testing in which the program under test
runs repeatedly with test inputs in order to discover bugs in the program.
Fuzzing can be categorized into two types based on their input generation
methodology~\cite{sutton:2007}: mutational fuzzing and generational fuzzing.
Mutational fuzzing~\cite{aflfuzz, woo:ccs:2013, rebert:usec:2014,
cha:oakland:2015} alters given seeds to generate new test inputs, whereas
generational fuzzing~\cite{funfuzz, holler:usec:2012, han:ccs:2017,
han:ndss:2019} produces tests based on an input model, such as a grammar.

Since JS code is highly structured, randomly generated test inputs are likely
to be rejected by JS engines. Therefore, it is common for JS engine fuzzers to
employ a generational approach.
One notable example is \jsfunfuzz, a seminal JS engine fuzzer~\cite{
jsfunfuzz:blog, funfuzz}. It starts with a start symbol defined in
a JS grammar and selects the next potential production in a random fashion until
there are no remaining non-terminal symbols. \alchemist~\cite{han:ndss:2019} is
another generational fuzzer that resort to the assembly constraints of its
building blocks called code bricks to produce semantically valid JS code.

Most other JS engine fuzzers use both mutational and generational approaches.
LangFuzz~\cite{holler:usec:2012}, GramFuzz~\cite{guo:2013}, and
IFuzzer~\cite{veggalam:esorics:2016} parse JS seeds with the JS grammar and
construct a pool of code fragments, where a code fragment is a subtree of an
AST. They combine code fragments in the pool to produce a new JS test input, but
they also mutate given seeds to generate test inputs.

Although it does not aim to find security vulnerabilities, TreeFuzz~\cite{patra:2016}
leverages a probabilistic context-free grammar (PCFG) to generate a test suite from
given seeds. Similarly, Skyfire~\cite{wang:oakland:2017} infers a probabilistic
context-sensitive grammar (PCSG) from given seeds and uses it to generate a
well-distributed set of seeds. Both approaches apply probabilistic language models
to generate JS testing inputs, but their design is too generic to find security
vulnerabilities in JS engines. Unlike previous approaches, \sysname is inspired
by a systematic study of CVEs, i.e., previous JS engine vulnerabilities, and
leverages an NNLM trained to learn syntactic and semantic commonalities between
JS regression test suites.


%% file: motive.tex
\section{Motivation}
\label{s:motive}

Can we find similarities between JS files that trigger security
vulnerabilities? We answer this question by conducting a quantitative study of
analyzing reported CVEs and corresponding proof of concept (PoC) exploits for
\chakra~\cite{chakra}.
We chose \chakra because its GitHub repository maintains well-documented commit
logs describing whether a specific CVE is patched by a commit. This helps us
identify which security vulnerability is related to a given PoC exploit and
which source lines are affected by the vulnerability.
Other JS engines, in contrast, have not provided an exact mapping between a
code commit and a CVE.

Note that collecting PoC exploits is not straightforward because CVE reports
typically do not carry any PoC exploits due to the potential risk of being
abused. We manually collected CVEs as well as their PoC code from exploitDB,
vulnerability blogs, and the \chakra GitHub repository. In total, we obtained 67
PoC exploits, each of which corresponds to a unique CVE. We further identified
50 of them where the corresponding vulnerabilities are fixed by a single commit.
This means that we can map each of the 50 vulnerabilities to a set of affected
source files.
The earliest and the latest vulnerabilities in the collected set were patched in
September 2016 and March 2018, respectively.
In total, 77 files were patched owing to these vulnerabilities.

We found that nine out of the 50 vulnerabilities (18\%) are related to the
\texttt{GlobOpt.cpp} file, which mainly implements the just-in-time (JIT)
compilation step.
Seven of them (14\%) have also contributed to patching the
\texttt{JavascriptArray.cpp} file.
Note that each file implements different functionalities of \chakra.
In other words, different JS engine vulnerabilities often arise from a common file
that implements the same functionalities, such as JIT optimization and JS arrays.
For example, a patch for CVE-2018-0776 forces a deep copy of an array when the
array is accessed via the function \texttt{arguments} property within a callee, thus
avoiding a type confusion vulnerability.
However, the patch was incomplete, still leaving other ways in which a shallow
copy of arrays could be caused. CVE-2018-0933 and CVE-2018-0934 were assigned
to those bugs.
Note that all the patches revised the \texttt{BoxStackInstance} function in the
\texttt{JavascriptArray.cpp} file.

Among the 77 patched files, 26 (33.8\%) files are patched at least twice due
to the reported CVEs.
These examples demonstrate that JS engine vulnerabilities often arise from files
that were patched for other bugs.
Considering that these patches are often checked with regression tests, mutating
an existing JS test may trigger a new vulnerability whose root cause lies in the
patched files that this test already covered.

\vspace{0.5em}
\noindent\fbox{%
  \parbox{0.98\columnwidth}{%
    \textbf{Observation 1.}
    JS engine vulnerabilities often arise from the same file patched for different bugs.
  }%
}
\vspace{0.4em}

We also measured the syntactic similarity between JS code from the PoC exploits
and \chakraNumRegressionTests JS files obtained from regression test suites
maintained by \chakra.
Note that a regression test suite consists of JS tests that trigger previously
patched bugs and check expected outcomes with adversarial test input.
In particular, we gathered the regression test files from the \chakra version
released in August 2016, which is one month ahead of the patching date of the
earliest vulnerability. Therefore, the regression test files were not affected
by any of the studied vulnerabilities.

\vspace{-1.0em}
\begin{figure}[ht]
\begin{lstlisting}[escapechar=$]
  var v0 = {};
  for (var v1 = 0; v1 < 5; v1++) {
      v0[v1] = v1 + 5; $\label{line:example}$
  }
\end{lstlisting}
\vspace{-1.0em}
\caption{Example of a normalized JS file.}
\label{lst:example}
\vspace{-0.5em}
\end{figure}

To measure the similarity, we normalized the identifiers in the regression test
files as well as the PoC exploits. Specifically, we renamed each identifier for
variables and functions to have a sequential number and a common prefix as
their name. We then parsed the normalized JS files down to ASTs.

We extracted a set of unit subtrees with a depth of one from each AST.
For a given AST, we extracted a unit subtree from each internal node. Thus,
the number of extracted unit subtrees becomes the number of AST internal
nodes. We call such a unit subtree a \textit{fragment}, as formally defined in
\S\ref{s:design}.
Note that the root node of each fragment is an internal node of the AST. It also
corresponds to a leaf node in another fragment, except the fragment with the
root node of the original AST.

\begin{figure}[!t]
\centering
\includegraphics[width=0.98\columnwidth]{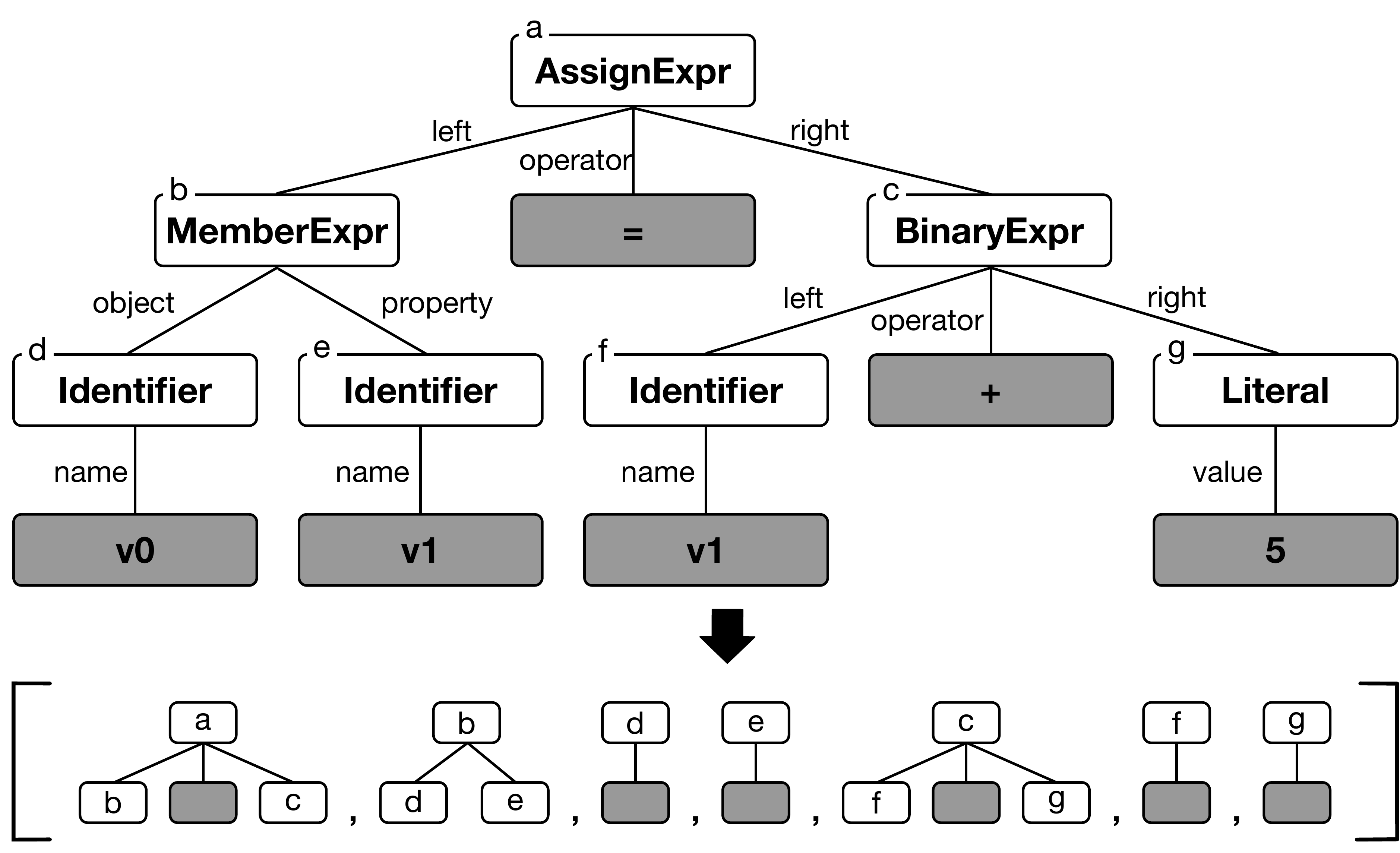}
\caption{Fragmentizing an AST from the example in Figure~\ref{lst:example}.}
\label{fig:frag-overview}
\vspace{-1.0em}
\end{figure}

Figure~\ref{fig:frag-overview} illustrates the fragmentation results for a JS
file listed in Figure~\ref{lst:example}. The upper side of the figure shows an
AST subtree obtained from the Esprima JS parser~\cite{esprima}. This subtree
corresponds to Line~\ref{line:example}. The bottom of the figure presents
fragments from this subtree.

\begin{figure}[tb]
\centering
  \includegraphics[width=1\linewidth]{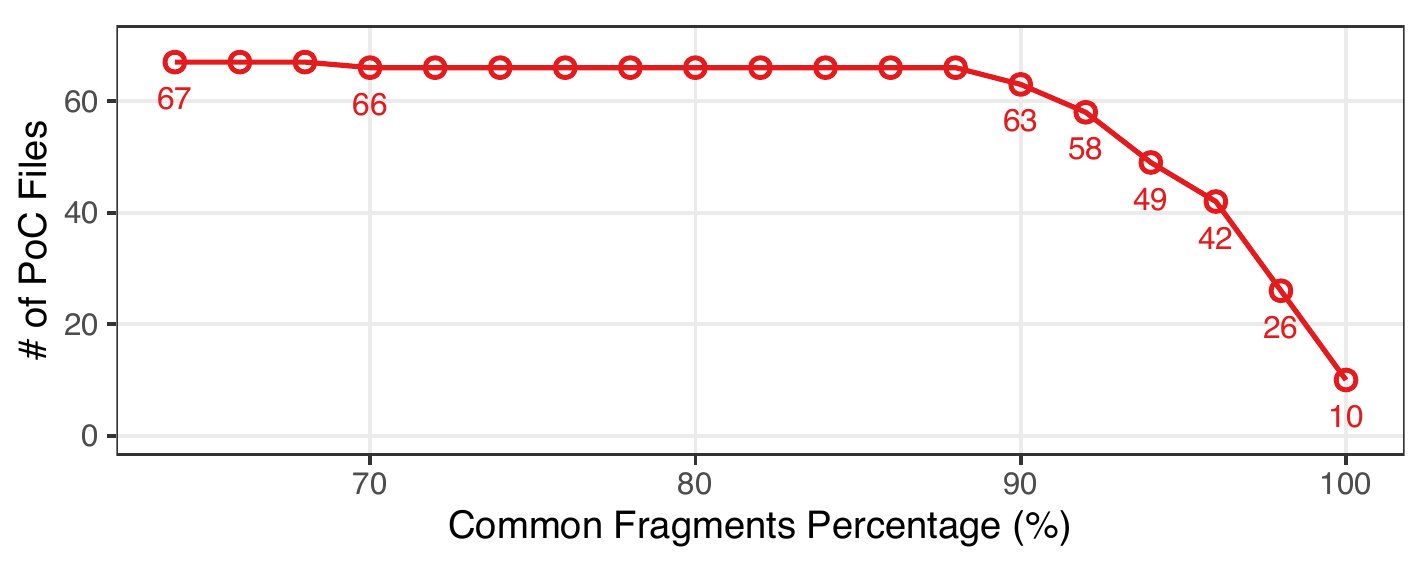}
\caption{
The number of all PoC files whose common fragment percentages are greater than
varying percentages.
}
\label{fig:motive}
\vspace{-1.0em}
\end{figure}

We also divided each PoC that triggers a CVE into fragments and then counted
how many fragments existed in the regression test suites.
Figure~\ref{fig:motive} depicts the number of PoC files whose common fragment
percentage is over each percentage threshold.
We found that all the fragments (100\%) from 10 PoC exploits already existed
in the regression test files. More than 96\% of the fragments in the 42 PoC exploits
and 90\% of the fragments in the 63 PoC exploits existed in the regression test as
well. On average, \chakraAvgOverlapRate of the fragments from the PoC exploits
were found in the regression test files.

\vspace{0.5em}
\noindent\fbox{%
    \parbox{0.98\columnwidth}{%
\textbf{Observation 2.} More than 95\% of the fragments syntactically
overlap between the regression tests and the PoC exploits.
    }%
}
\vspace{0.4em}

Both observations imply that it is likely to trigger a new security
vulnerability by assembling code fragments from existing regression test suites,
which is the primary motivation for this study, as we describe in
\S\ref{sec:mljs-overview}.

%% file: overview.tex
\section{Overview}
\label{sec:mljs-overview}

We present \sysname, an NNLM-driven fuzzer, which automatically finds bugs in
JS engines. Recall that the overall design of \sysname is driven by two
observations: (1) security bugs often arise from files that were previously
patched for different causes, and (2) the JS test code that triggers
security-related bugs heavily reuses AST fragments found in the existing
regression test sets.

We propose a novel fuzzing technique that captures these observations.
We train an NNLM to capture the syntactic and semantic relationships among
fragments from the regression test sets. When generating a new JS
test, \sysname mutates the AST of a given JS regression test. It replaces a
subtree of the AST with a new subtree, using the trained NNLM.
Thus, each generated test stems from a given regression test that checks
previously patched or buggy logic, thereby, capturing the first observation. At
the same time, it invokes functionalities in different execution contexts by
assembling existing fragments under the guidance of the NNLM, which addresses
the second observation.

\begin{figure}[ht]
\centering
\includegraphics[width=0.98\columnwidth]{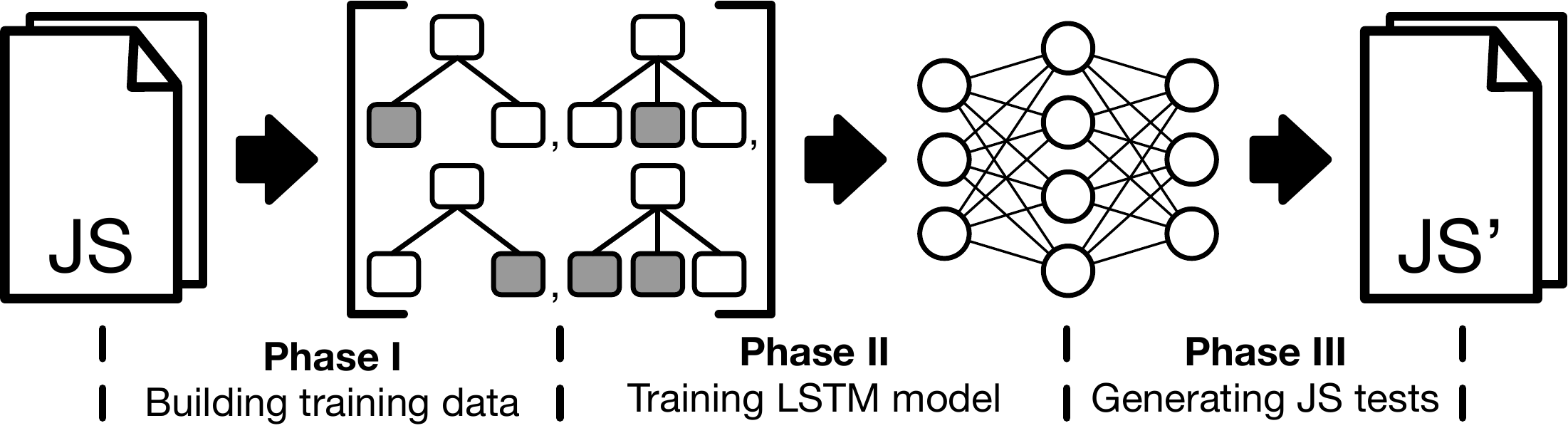}
\caption{Overview of \sysname.}
\label{fig:detailed-overview}
\vspace{-0.5em}
\end{figure}

Figure~\ref{fig:detailed-overview} shows the overall workflow of \sysname.
Phase I prepares the training instances from given regression test suites. Each
training instance is a sequence of AST unit subtrees, called
\textit{fragments}. Phase II trains an NNLM that learns compositional
relationships among fragments. These two phases are one-time setup
procedures. Phase III generates JS tests by leveraging the trained model.

Phase I begins with a given training set of JS regression test files. It parses
each JS file into an AST and normalizes identifiers that appeared in the AST to
deduplicate function and variable names.
Figure~\ref{lst:example} shows a normalized JS file example. Each appeared
variable name is changed into a common name, such as \texttt{v0} or \texttt{v1}.
From a normalized AST tree, Phase I then extracts multiple unit subtrees, each
of which is called a \textit{fragment}. For each node in the AST, \sysname
recursively slices a unit subtree of depth one. Each of the sliced subtrees
becomes a fragment of the AST.
It then emits the sequence of these fragments, produced by the pre-order traversal
of their root nodes in the normalized AST tree.

Phase II trains the NNLM given a set of fragment sequences.
From a given fragment sequence of an arbitrary length, we design the NNLM to
suggest the next fragments, which are likely to appear after this fragment
sequence. This framing is a key contribution of this paper.
Note that it is not straightforward to model the inherent structural
relationships of an AST in such a way that a language model can learn.
By leveraging the fragments encapsulating the structural relationships of ASTs,
we encode a given AST into fragment sequences. Considering that a vast volume
of natural language NNLMs have been trained upon word sequences, this fragment
sequencing eases the application of existing prevailing NNLMs for generating JS
tests.

Here, the objective is to train the NNLM to learn compositional relationships
among fragments so that the JS test code generated from the trained model
reflects the syntax and semantics of the given training set, which is the
regression testing set of JS engines.

Phase III generates a new JS test by leveraging the trained model and the AST
of a regression test. Given a set of ASTs from regression test suites, it
randomly picks a seed AST. Then, it randomly selects a subtree for
\sysname to replace. When generating a new subtree, \sysname considers a
\textit{context}, the sequence of all fragments that precedes the selected
subtree. \sysname iteratively appends fragments from the root node of the
selected subtree while considering its context.

Because the current AST is assembled from fragments, it is expected that some
variables and function identifiers in the AST nodes are used without proper
declarations. \sysname, thus, resolves possible reference errors by renaming
them with the declared identifiers.
Finally, \sysname checks the generated test and reports a bug if the code
crashes the target JS engine.

\noindent\textbf{Other model guided approaches.}
Previous studies presented language models, which can predict the lexical code
tokens in source code. Such framing of language models has been vastly studied
while addressing code completion problems~\cite{
  tu:fse:2014,nguyen:fse:2013}. However, the generation of an executable test is
more challenging than the code completion problem that predicts a limited number
of semantically correct lexical tokens. To our knowledge, the PDF fuzzer
proposed by Singh~\etal~\cite{godefroid:ase:2017} is the first system that
employs a character-level RNN model to generate PDF tests.  We evaluated whether
our fragment-based approach performs better than the character-level RNN model
approach in finding JS engine bugs (see \S\ref{ss:model-effectiveness}).

%% file: design.tex
\section{Design}
\label{s:design}

The design goal of \sysname is to generate JS test inputs that can trigger
security vulnerabilities in JS engines, which (1) reflect the syntactic
and semantic patterns of a given JS training set, and (2) trigger no reference errors.

It is a technical challenge to frame the problem of teaching a language model
the semantic and syntactic patterns  of training code. We address this
challenge by abstracting the hierarchical structure by AST subtrees, which we
refer to as fragments. We then enable the language model to learn the
compositional relationships between fragments.

We propose a novel code generation algorithm that leverages a trained
language model. We harness an existing JS code that is already designed to
trigger JS engine defects. \sysname alters this existing JS code by
replacing one of its AST subtrees with a new subtree that the
trained language model generates.
Thus, \sysname is capable of generating a new JS test, semantically similar
to the regression test case that triggers a previously reported bug. We expect
that this new JS test triggers a new bug in a different execution context.

\subsection{Phase I: Building Training Data of Fragment Sequences}
\label{ss:phase1}

Phase I prepares training instances using a given training set. It conducts
\emph{parsing} and \emph{fragmentation}.

\subsubsection{Parsing and Normalizing}
\label{ss:parsing}

Phase I builds an AST by parsing each JS file in a training set and normalizes
the parsed AST.
Because the training set includes a variety of JS files from various developers,
identifier naming practices are not necessarily consistent. Thus, it is natural
that the training files have diverse variable and function names across
different JS files.
Consider two JS files that contain a JS statement  \texttt{var b = a + 1} and
\texttt{var c = d + 1}, respectively. Both have the same AST structure and
semantics, but different identifiers.

This pattern increases the size of unnecessary vocabulary for a language model
to learn, rendering the model evaluation expensive as it requires more
training instances. To have concise ASTs with
consistent identifier names, we rename all the variable and function identifiers
in the ASTs.

Specifically, for each declared variable identifier, we assign a sequential
number in the order of their appearance in a given AST. We then replace each
variable name with a new name that combines a common prefix and its sequential
number, such as \texttt{v0} and \texttt{v1}. We also apply the same procedure to
function identifiers, e.g., \texttt{f0} and \texttt{f1}. We deliberately exclude
language-specific built-in functions and engine objects from the normalization
step as normalizing them affects the semantics of the original AST. For an
\texttt{eval} function that dynamically evaluates a given string as the JS code,
we first extract the argument string of the \texttt{eval} function and strip it
out as the JS code when the argument is a constant string. Subsequently, we
normalize identifiers in the JS code stripped out from the \texttt{eval}
argument.

As our training set is derived from regression tests of JS engines, JS files
in the set make heavy use of predefined functions for testing purposes.
Therefore, we manually identified such vendor-provided testing functions and
ignore them during the normalization step. That is, we treated common testing
functions provided by each JS engine vendor as a built-in function and excluded
them from normalization.

\subsubsection{Fragmentation}
\label{ss:frag}

\sysname slices each normalized AST into a set of subtrees while ensuring that
the depth of each subtree is one. We call such a unit subtree as a \textit{
fragment}.

We represent an AST $T$ with a triple $(\nodes, \edges, \rnode)$, where \nodes
is the set of nodes in $T$, \edges is the set of edges in $T$, and \rnode is the
root node of $T$.
We denote the immediate children of a given AST node $n_i$ by
$\children\left(n_i\right)$, where $n_i$ is a node in $\nodes$.
Then, we define a subtree of $T$ where the root node of the subtree is $n_i$.
When there is such a subtree with a depth of one, we call it a
\textit{fragment}. We now formally define it as follows.

\begin{defn}[Fragment]\label{def:dfrag}
  A fragment of $T = (\nodes, \edges, \rnode)$ is a subtree
  $T_i = (\nodes_i, \edges_i, n_i)$, where
  \begin{itemize}
    \item $n_i \in \nodes$ s.t. $\children\left(n_i\right) \neq \emptyset$.
    \item $\nodes_i = \{n_i\} \bigcup \children(n_i)$.
    \item $\edges_i = \left\{(n_i, n^\prime) \mid  n^\prime = \children(n_i) \right\}$.
  \end{itemize}
\end{defn}

Intuitively, a fragment whose root node is $n_i$ contains its children and their
tree edges.
Note that each fragment inherently captures an exercised production rule of
the JS language grammar employed to parse the AST.
We also define the $type$ of a fragment as the non-terminal symbol of its
root node $n_i$.
For instance, the first fragment at the bottom side of
Figure~\ref{fig:frag-overview} corresponds to the assignment expression
statement in Line~\ref{line:example} of Figure~\ref{lst:example}. The fragment
possesses four nodes whose root node is the non-terminal symbol of an
\texttt{AssignmentExpression}, which becomes the $type$ of this fragment.

\sysname then generates a sequence of fragments by performing the pre-order
traversal on the AST. When visiting each node in the AST, it emits the fragment
whose root is the visited node.  The purpose of the pre-order sequencing is to
sort fragments by the order of their appearance in the original AST.
For example, the bottom side of Figure~\ref{fig:frag-overview} shows the
sequence of seven fragments obtained from the AST subtree in the figure.

We model the compositional relationships between fragments as a pre-order
sequencing of fragments so that an NNLM can predict the next fragment to use
based on the fragments appearing syntactically ahead.
In summary, Phase I outputs the list of fragment sequences from the training
set of normalized ASTs.

\subsection{Phase II: Training an LSTM Model}
\label{ss:phase2}

All distinct fragments become our vocabulary for the NNLM to be trained. Before
training, we label the fragments whose frequency is less than five in the
training set as out-of-vocabulary (OoV). This is a standard procedure for
building a language model to prune insignificant words~\cite{hindle:icse:2012,
nguyen:fse:2013, raychev:pldi:2014}.

Each fragment sequence represents a JS file in the training set. This sequence
becomes a training instance.
We build a statistical language model from training instances so that the model
can predict the next fragment based on all of its preceding fragments, which is
considered as a \textit{context}. This way, the model considers each fragment
as a lexicon, and thereby, suggests the next probable fragments based on the
current context.

\noindent\textbf{Training objectives.}
The overall objective is to model a function $f : X \rightarrow Y$ such that
$y \in Y$ is a probability distribution for the next fragment $fr_{t+1}$,
given a fragment sequence $x = [fr_0, fr_1, ..., fr_t] \in X$, where $fr_i$
denotes each fragment at time step $i$.
Given $x$, the model is trained to (1) predict the correct next fragment with
the largest probability output and (2) prioritize fragments that share the same
$type$ with the true fragment over other types of fragments.
Note that this training objective accords with our code generation algorithm in
that \sysname randomly selects the next fragment of a given type from the top
$k$ suggestions (see \S\ref{ss:phase3}).

\begin{figure}[!t]
\centering
\includegraphics[width=\columnwidth]{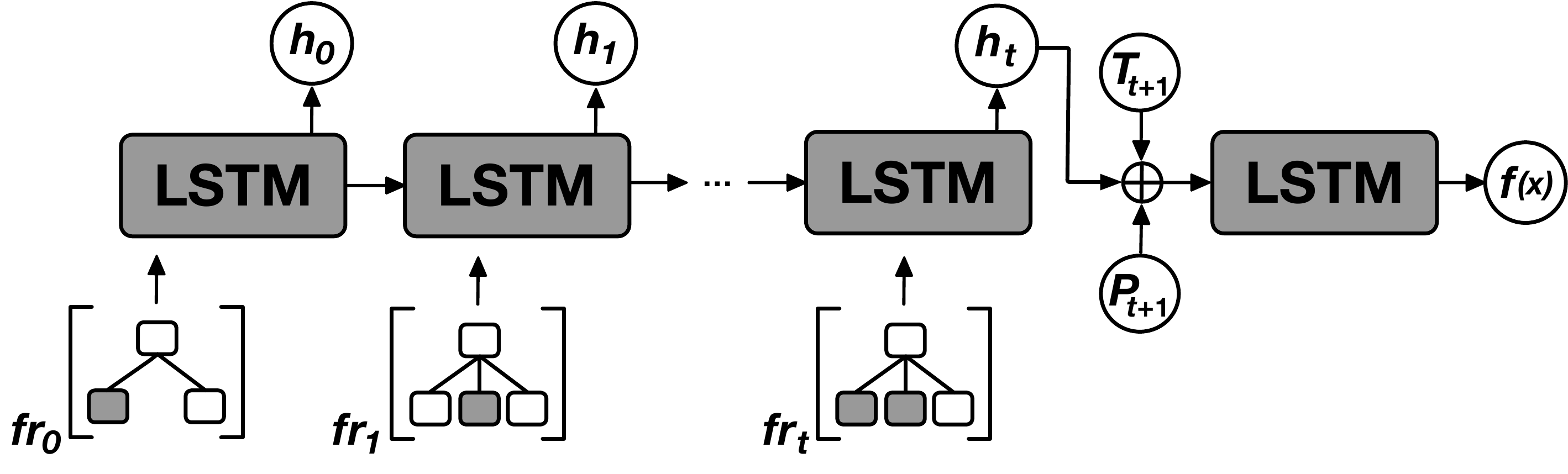}
\caption{Architecture of \sysname LSTM model. $\oplus$ in the figure denotes
a concatenation.}
\label{fig:model-architecture}
\vspace{-1.0em}
\end{figure}

\noindent\textbf{LSTM.} To implement such a statistical language
model, we take advantage of the LSTM model~\cite{hochreiter:nc:1997}.
Figure~\ref{fig:model-architecture} depicts the architecture of \sysname LSTM
model. Our model consists of one projection, one LSTM, and one output layers.
The projection layer is an embedding layer for the vocabulary where each fragment has a dimension size of 32.
When $fr_t$ is passed into the model, it is converted into a vector, called
$embedding$, after passing the projection layer.

Then, the $embedding$ vector becomes one of the inputs for the LSTM layer with a
hidden state size of 32.
At each time step, the LSTM cell takes three inputs: (1) a hidden state
$h_{t-1}$ and (2) a cell state $c_{t-1}$ from the previous time step; and (3)
the embedding of a new input fragment.
This architecture enables the model to predict the next fragment based on the
cumulative history of preceding fragments.
In other words, the LSTM model is not limited to considering a fixed number of
preceding fragments, which is an advantage of using an RNN model.

The output of the LSTM layer $h_t$ is then concatenated with two other
vectors: (1) the type embedding $T_{t+1}$ of the next fragment, and (2) the
fragment embedding $P_{t+1}$ of the parent fragment of the next fragment in
its AST.
The concatenated vector is now fed into the final output layer and it outputs a
vector $f(x)$ of vocabulary size to predict the next fragment.

\noindent\textbf{Loss function.}
To address our training objectives, we defined a new loss function that rewards
the model to locate type-relevant fragments in its top suggestions.
The LSTM model is trained to minimize the following empirical loss over the
training set $(x,y) \in D$.
\begin{equation}
\small
  \begin{aligned}
  \label{eq:loss1}
    g(x) &= \textup{softmax}(f(x)) \\
    L_{D}(f) &= \frac{1}{|D|}\sum\limits_{(x,y) \in D} l_1(g(x),y) + l_2(g(x),y)
  \end{aligned}
\end{equation}
As shown in Equation~\ref{eq:loss1}, the loss function has two terms: $l_1$ and
$l_2$. Note that these terms are designed to achieve our two training
objectives, respectively.

\begin{equation}
\small
  \begin{aligned}
  \label{eq:loss2}
   l_1(g(x),y) &= - \sum\limits_{i=1}^{N}{y_i}\log g(x)_i \\
   l_2(g(x),y) &= \sum\limits_{i \in \textup{top}(n)}{g(x)_i} -
                 \sum\limits_{j \in \textup{type}(y)}{g(x)_j},
  \end{aligned}
\end{equation}

Equation~\ref{eq:loss2} describes each term in detail.
In the equation, $n$ denotes the number of fragments whose types are same as
that of the true fragment. $\textup{top}(n)$ and $\textup{type}(y)$ indicate
functions that return the indices of top $n$ fragments and fragments of the true
type, respectively.

$l_1$ is a cross entropy loss function, which has been used for common natural
language models~\cite{mikolov:acisca:2010, jozefowicz:arxiv:2016}.
$l_2$ is employed for rewarding the model to prioritize fragments that have the
same type as the true fragment. We formally define $l_2$ as a \textbf{type
error}.
It is a gap between two values: the sum of the model output probabilities
corresponding to (1) top $n$ fragments and (2) fragments of the true type.

By reducing the sum of $l_1$ and $l_2$ while training, the model achieves our
training objectives. Intuitively, the LSTM model is trained not only to predict
the correct fragment for a given context, but also to locate fragments whose
types are same as the correct fragment in its top suggestions.

The fundamental difference of \sysname from previous approaches that use
probabilistic language models~\cite{wang:oakland:2017, patra:2016} lies in the
use of fragments.
To generate JS code, TreeFuzz~\cite{patra:2016} and
SkyFire~\cite{wang:oakland:2017} use a PCFG and PCSG to choose the next AST
production rule from a given AST node, respectively. SkyFire defines its context
to be sibling and parent nodes from a given AST. It picks an AST production rule
that is less frequent in the training set. In contrast, \sysname selects a
fragment based on the list of fragments, not AST nodes. Therefore, \sysname is
capable of capturing the global composition relationships among code fragments
to select the next code fragment.
Furthermore, \sysname preserves the semantics in the training set by
slicing the AST nodes into fragments, which is used as a lexicon for
generating JS code. We frame the problem of training a language model to
leverage fragments and their sequences, which makes \sysname compatible with
prevalent statistical language models.

\subsection{Phase III: Generating JS Tests}
\label{ss:phase3}
Given a set of ASTs from regression tests and the LSTM model, Phase III first
mutates a randomly selected seed AST by leveraging the LSTM model. Then, it
resolves reference errors in the skeleton AST.

Algorithm~\ref{algo:generate} describes our code generation algorithm.
The \texttt{MutateAST} function takes two configurable parameters from users.
\begin{enumerate}[noitemsep,leftmargin=1.6cm,labelsep=0.6cm]
  \item[\fmax] The maximum number of fragments to append. This parameter
  controls the maximum number of fragments that a newly composed subtree can
  have.
  \item[\topk] The number of candidate fragments. \sysname randomly selects
  the next fragment from suggestions of the \topk largest probabilities at each
  iteration.
\end{enumerate}
After several exploratory experiments, we observed that bloated ASTs are more
likely to have syntactical and semantic errors. We also observed that
the accuracy of the model decreases as the size of an AST increases.
That is, as the size of AST increases, \sysname has a higher chance of failures
in generating valid JS tests.
We thus capped the maximum number of fragment insertions with \fmax and
empirically chose its default value to be 100.
For \topk, we elaborate on its role and effects in detail in
\S\ref{ss:topk-test}.

\subsubsection{Mutating a Seed AST}

The \texttt{MutateAST} function takes in a set of ASTs from regression tests,
the trained LSTM model, and the two parameters.
It then begins by randomly selecting a seed AST from the given set
(Line~\ref{algo:line:seed}).
From the seed AST, it removes one randomly selected subtree
(Line~\ref{algo:line:rm}). Note that the pruned AST becomes a base for the new
JS test.
Finally, it composes a new subtree by leveraging the LSTM model
(Lines~\ref{algo:line:start}-\ref{algo:line:end}) and returns the newly composed
AST.

\input{algo/generate}

After selecting a seed AST in Line~\ref{algo:line:seed}, we randomly prune one
subtree from the AST by invoking the \texttt{RemoveSubtree} function.
The function returns a pruned AST and the initial $context$ for the LSTM model,
which is a fragment sequence up to the fragment where \sysname should start to
generate a new subtree. This step makes a room to compose new code.

In the \texttt{while} loop in Lines~\ref{algo:line:start}-\ref{algo:line:end},
the \texttt{MutateAST} function now iteratively appends fragments to the AST
at most \fmax times by leveraging the LSTM model.
The loop starts by selecting the next fragment via the \texttt{PickNextFrag}
function in Line~\ref{algo:line:pick}.
The \texttt{PickNextFrag} function first queries the LSTM model to retrieve the
\topk suggestions.
From the suggestions, the function repeats random selections until the chosen
fragment indeed has a correct type required for the next fragment.
If all the suggested fragments do not have the required type, the
\texttt{MutateAST} function stops here and abandon the AST. Otherwise, it
continues to append the chosen fragment by invoking the \texttt{AppendFrag}
function.

The \texttt{AppendFrag} function traverses the AST in the pre-order to find
where to append the fragment.
Note that this process is exactly the opposite process of an AST fragmentation
in \S\ref{ss:frag}.
Because we use a consistent traversal order in Phase I and III, we can easily
find whether the current node is where the next fragment should be appended.
Lines~\ref{algo:appd:start}-\ref{algo:appd:end} summarize how the function
determines it.
The function tests whether the current node is a non-terminal that does not have
any children.
If the condition meets, it appends the fragment to the current node and returns.
If not, it iteratively invokes itself over the children of the node for the
pre-order traversal.

Note that the presence of  a non-terminal node with no children indicates that
the fragment assembly of the AST is still in progress.
The \texttt{IsASTBroken} function checks whether the AST still holds such nodes.
If so, it keeps appending the fragments. Otherwise, the \texttt{MutateAST}
function returns the composed skeleton AST.

We emphasize that our code generation technique based on code fragments greatly
simplifies factors that a language model should learn in order to generate an
AST.
TreeFuzz~\cite{patra:2016} allows a model to learn fine-grained relationships
among edges, nodes, and predecessors in an AST. Their approach requires to
produce multiple models each of which covers a specific property that the model
should learn.
This, though, brings the unfortunate side-effects of managing multiple models
and deciding priorities in generating an AST when the predictions from different
models conflict with each other.
On the other hand, our approach abstracts such relationships as fragments, which
becomes building blocks for generating AST.
The model only learns the compositional relationships between such blocks, which
makes training and managing a language model simple.

\subsubsection{Resolving Reference Errors}\label{ss:error-resolution}
Phase III resolves the reference errors from the generated AST, which appear
when there is a reference to an undeclared identifier.
It is natural for the generated AST to have reference errors since
we assembled fragments that are used in different contexts across various training
files.
The reference error resolution step is designed to increase the chance of
triggering bugs by making a target JS engine fully exercise the semantics of a
generated testing code. The previous approaches~\cite{holler:usec:2012,
guo:2013, veggalam:esorics:2016} reuse existing AST subtrees and attach them
into a new AST, which naturally causes reference errors. However, they
overlooked this reference error resolution step without addressing a principled
solution.

We propose a systematic way of resolving reference errors, which often
accompany type errors.
Specifically, we take into account both (1) statically inferred JS types and
(2) the scopes of declared identifiers. \sysname harnesses these two factors
to generate JS test cases with fewer reference errors in the run time.

There are three technical challenges that make resolving reference errors
difficult. (1) In JS, variables and functions can be referenced without
their preceding declarations due to hoisting~\cite{hoisting}. Hoisting places
the declarations of identifiers at the top of the current scope in its
execution context; (2) It is difficult to statically infer the precise type of
each variable without executing the JS code because of no-strict type checking
and dynamically changing types; and (3) Each variable has its own scope so that
referencing a live variable is essential to resolve reference errors.

To address these challenges, \sysname prepares a scope for each AST node that
corresponds to a new block body. \sysname then starts traversing from these
nodes and fills the scope with declared identifiers including hoistable
declarations. Each declared identifier in its scope holds the \texttt{undefined}
type at the beginning.

When \sysname encounters an assignment expression in its traversal, it
statically infers the type of its right-hand expression via its AST node
type and assigns the inferred type to its left-hand variable. \sysname covers
the following statically inferred types: \texttt{array, boolean, function, null,
number, object, regex, string, undefined}, and \texttt{unknown}. Each scope has
an identifier map whose key is a declared identifier and value is an inferred
type of the declared identifier.

To resolve reference errors, \sysname  identifies an undeclared variable while
traversing each AST node and then infers the type of this undeclared variable based on its usage.
A property or member method reference of such an
undeclared variable hints to \sysname to infer the type of the undeclared
variable. For instance, the \texttt{length} property reference of an undeclared
variable assigns the \texttt{string} type to the undeclared variable.
From this inferred type,  \sysname replaces the undeclared identifier with a
declared identifier when its corresponding type in the identifier map is same
as the inferred type.
If the inferred type of an undeclared variable is \texttt{unknown}, it
ignores the type and randomly picks one from the list of declared identifiers.
For all predefined and built-in identifiers, \sysname treats them as declared
identifiers.
%


%% file: algo/generate.tex
\begin{algorithm}[t] {
  \footnotesize
  \DontPrintSemicolon
  \SetNoFillComment
  \SetKwSty{algokeywordsty}
  \SetFuncSty{algofuncsty}
  \SetDataSty{algodatasty}
  \SetArgSty{algoargsty}
  \SetKwInOut{Input}{Input}
  \SetKwInOut{Output}{Output}
  \SetKwFunction{generate}{\texttt{Generate}}
  \SetKwFunction{mutateAST}{\texttt{MutateAST}}
  \SetKwFunction{appendFrag}{\texttt{AppendFrag}}
  \SetKwFunction{convert}{\textsc{convert}}
  \SetKwFunction{resolve}{\texttt{Resolve}}
  \SetKwFunction{rand}{\textsc{rand}}
  \SetKwFunction{Index}{\textsc{Index}}
  \SetKwFunction{randPick}{\texttt{RandPick}}
  \SetKwFunction{update}{\textsc{update}}
  \SetKwFunction{getNext}{\textsc{getNext}}
  \SetKwFunction{getTopKCandidates}{\texttt{GetTopK}}
  \SetKwFunction{removeSubtree}{\texttt{RemoveSubtree}}
  \SetKwFunction{pickNextFrag}{\texttt{PickNextFrag}}
  \SetKwFunction{pickRandomSeed}{\texttt{PickRandomSeed}}
  \SetKwFunction{isASTBroken}{\texttt{IsASTBroken}}
  \SetKwFunction{isNonTerminal}{\texttt{IsNonTerminal}}
  \SetKwFunction{findNode}{\texttt{FindNonTerminal}}
  \SetKwFunction{append}{\ensuremath{append}}
  \SetKwFunction{update}{\ensuremath{Update}}
  \SetKwFunction{child}{\ensuremath{child}}
  \SetKwFunction{dequeue}{\ensuremath{dequeue}}
  \SetKwData{null}{\ensuremath{\varnothing}}
  \SetKwData{model}{\ensuremath{model}}
  \SetKwData{seedcxt}{\ensuremath{context_{seed}}}
  \SetKwData{ctx}{\ensuremath{context}}
  \SetKwData{frag}{\ensuremath{fragment\xspace}}
  \SetKwData{fragi}{\ensuremath{fragment\ensuremath{_i}\xspace}}
  \SetKwData{root}{\ensuremath{n_0}}
  \SetKwData{fraglist}{\texttt{fraglist}}
  \SetKwData{fragpool}{\ensuremath{\mathbb{F}}}
  \SetKwData{fragseq}{\ensuremath{frag\_seq}}
  \SetKwData{nextfrag}{\ensuremath{next\_frag}}
  \SetKwData{ftree}{\texttt{FT}}
  \SetKwData{newFTree}{$\ftree^\prime$}
  \SetKwData{newCtx}{$\ctx^\prime$}
  \SetKwData{AST}{\ensuremath{AST}}
  \SetKwData{c}{\ensuremath{count}}
  \SetKwData{fmax}{\ensuremath{f_{max}}}
  \SetKwData{topk}{\ensuremath{k_{top}}}
  \SetKwData{vocab}{\ensuremath{vocab}}
  \SetKwData{node}{\ensuremath{node}}
  \SetKwData{nodetocut}{\ensuremath{node_{cut}}}
  \SetKwData{fragPool}{\texttt{Pool}}
  \SetKwData{fLeaf}{\texttt{leaf}}
  \SetKwData{newLeaf}{$\fLeaf^\prime$}
  \SetKwData{fLeafs}{\texttt{Leafs}}
  \SetKwProg{Fn}{function}{}{}
  \SetKwData{children}{\ensuremath{\mathbb{C}}}
  \SetKwData{seed}{\ensuremath{\mathbb{T}}}
  \SetKw{break}{break}

  \Input{
    A set of ASTs from regression tests (\seed).
    \\\ The LSTM model trained on fragments (\model).
    \\\ The max number of fragments to append (\fmax).
    \\\ The number of candidate fragments (\topk).
  }

  \Output{
    A newly composed AST.
  }

  \Fn{\mutateAST{\seed, \model, \fmax, \topk}}{
    $\root \gets \pickRandomSeed{\seed}$\; \label{algo:line:seed}
    $\root,\ \ctx \gets \removeSubtree{$\root$}$\; \label{algo:line:rm}
    $\c \gets 0$\; \label{algo:line:start}
    \While{$\c \le \fmax$} {
      $\nextfrag \gets \pickNextFrag{\model, \topk, \ctx}$\; \label{algo:line:pick}
      \If {$\nextfrag = \null$}{
        \Return
      }
      $\root \gets \appendFrag{$\root$, \nextfrag}$\; \label{algo:line:appd}
      \If {not $\isASTBroken{\root}$}{
        \break
      }
      $\ctx.\append(\nextfrag)$\;
      $\c \gets \c + 1$\;
    } \label{algo:line:end}
    \Return{$n_0$}\;
  }

  \Fn{\appendFrag{\node, \nextfrag}}{
    $\children \gets \node.\child()$ \tcc*{Get direct child nodes.} \label{algo:appd:start}
    \If{\isNonTerminal{\node} $\wedge$ $\children = \null$}{
      $\node \gets \nextfrag$\; 
      \Return\; \label{algo:appd:end}
    }
    \For{$c \in \children$}{
      \appendFrag{$c, \fragseq$}\;
    }
  }
}
  \caption{Mutating a seed AST}
  \label{algo:generate}
\end{algorithm}

%% file: implementation.tex
\section{Implementation}

We implemented \sysname with 3K+ LoC in Python and JS. We used Esprima
4.0~\cite{esprima} and Escodegen 1.9.1~\cite{escodegen} for parsing and printing
JS code, respectively. As both libraries work in the Node.js environment, we
implemented an inter-process pipe channel between our fuzzer in Python and the
libraries.

We implemented the LSTM models with PyTorch 1.0.0~\cite{pytorch}, using the L2
regularization technique with a parameter of 0.0001. The stochastic gradient
descent with a momentum factor of 0.9 served as an optimizer.

We leveraged the Python subprocess module to execute JS engines and obtain their
termination signals. We only considered JS test cases that crash with
\texttt{SIGILL} and \texttt{SIGSEGV} meaningful because crashes with other
termination signals are usually intended ones by developers.

To support open science and further research, we publish \sysname at
\url{https://github.com/WSP-LAB/Montage}.

%% file: experiment.tex
\section{Evaluation}
\label{sec:eval}

We evaluated \sysname in several experimental settings.
The goal is to measure the efficacy of \sysname in finding JS engine bugs,
as well as to demonstrate the necessity of an NNLM in finding bugs. We first
describe the dataset that we used and the experimental environment.
Then, we demonstrate (1) how good a trained \sysname NNLM is in predicting
correct fragments (\S\ref{ss:model-validity}),
(2) how we set a \topk parameter for efficient fuzzing
(\S\ref{ss:topk-test}),
(3) how many different bugs \sysname discovers, which other existing fuzzers
are unable to find (\S\ref{ss:evalcmp}),
and (4) how much the model contributes to \sysname finding bugs and generating
valid JS tests (\S\ref{ss:model-effectiveness}).
We conclude the evaluation with field tests on the latest JS engines
(\S\ref{ss:evalfield}). We also discuss case studies of discovered bugs
(\S\ref{ss:casestudy}).

\subsection{Experimental Setup}\label{ss:evalenv}
We conducted experiments on two machines running 64-bit Ubuntu 18.04 LTS with two
Intel E5-2699 v4 (2.2 GHz) CPUs (88 cores), eight GTX Titan XP DDR5X GPUs, and
512 GB of main memory.

\noindent\textbf{Target JS engine.} The \chakra GitHub repository has managed
the patches for all the reported CVEs by the commit messages since 2016. That
is, we can identify the patched version of \chakra for each known CVE and have
ground truth that tells whether found crashes correspond to one of the known
CVEs~\cite{chen:pldi:2013}. Therefore, we chose an old version of \chakra as our
target JS engine. We specifically performed experiments on \chakra 1.4.1, which
was the first stable version after January 31, 2017.

\noindent\textbf{Data.} Our dataset is based on the regression test sets of
Test262~\cite{test262} and the four major JS engine repositories at the version
of January 31, 2017: \chakra, \jsc, \moz, and \veight. We excluded test files
that \chakra failed to execute because of their engine-specific syntax and
built-in objects. We did not take into account files larger than 30 KB
because large files considerably increase the number of unique fragments with
low frequency. In total, we collected 1.7M LoC of 33,486 unique JS files.

\noindent\textbf{Temporal relationships.} \sysname only used the regression test
files committed before January 31, 2017, and performed fuzz testing campaigns on
the first stable version after January 31, 2017.
Thus, the bugs that regression tests in the training dataset check and the bugs
that \sysname is able to find are disjoint. We further confirmed that all CVEs
that \sysname found were patched after January 31, 2017.

\noindent\textbf{Fragments.} From the dataset, we first fragmented ASTs to
collect 134,160 unique fragments in total. On average, each training instance
consisted of 118 fragments. After replacing less frequent fragments with OoVs,
they were reduced to 14,518 vocabularies. Note that most replaced fragments were
string literals, e.g., bug summaries, or log messages.

\noindent\textbf{Bug ground truth.} Once \sysname found a JS test triggering a
bug, we ran the test against every patched version of \chakra to confirm whether
the found bug matches one of the reported CVEs. This methodology is well-aligned
with that of Klees~\etal~\cite{klees:ccs:2018}, which suggests counting distinct bugs
using ground truth. When there is no match, the uniqueness of a crash was
determined by its instruction pointer address without address space layout
randomization (ASLR). We chose this conservative setting to avoid overcounting
the number of found bugs~\cite{molnar:usec:2009}.

\subsection{Evaluation of the LSTM Model}
\label{ss:model-validity}

To train and evaluate the LSTM model of \sysname, we performed a 10-fold
cross-validation on our dataset. We first randomly selected JS files for the
test set, which accounted for 10\% of the entire dataset. We then randomly split
the remaining files into 10 groups. We repeated holding out one group for the
validation set and taking the rest of them for the training set for 10 times.

\begin{figure}[!t]

\begin{subfigure}{0.49\columnwidth}
\centering
  \includegraphics[scale=0.6]{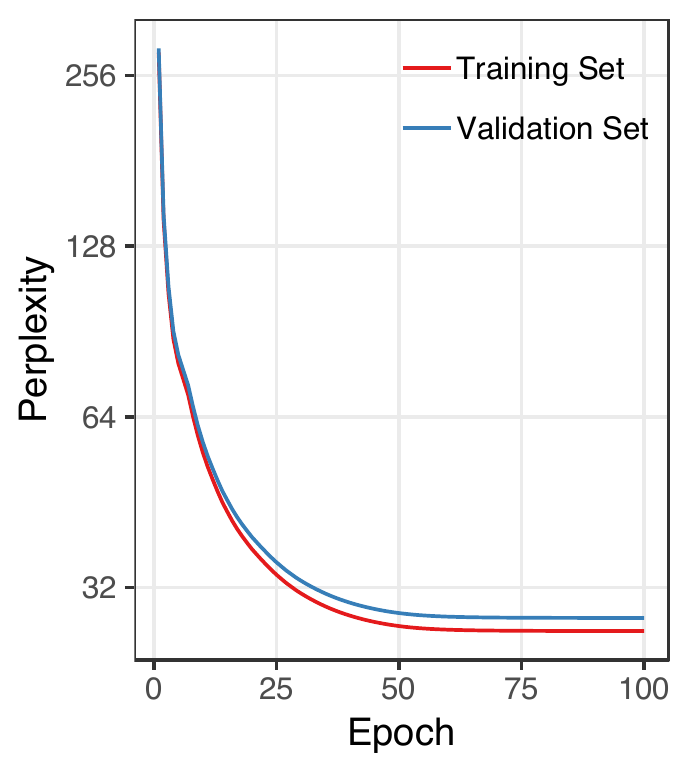}
\caption{Perplexity of the model.}
\label{fig:model-pplx}
\end{subfigure}
\begin{subfigure}{0.49\columnwidth}
\centering
  \includegraphics[scale=0.6]{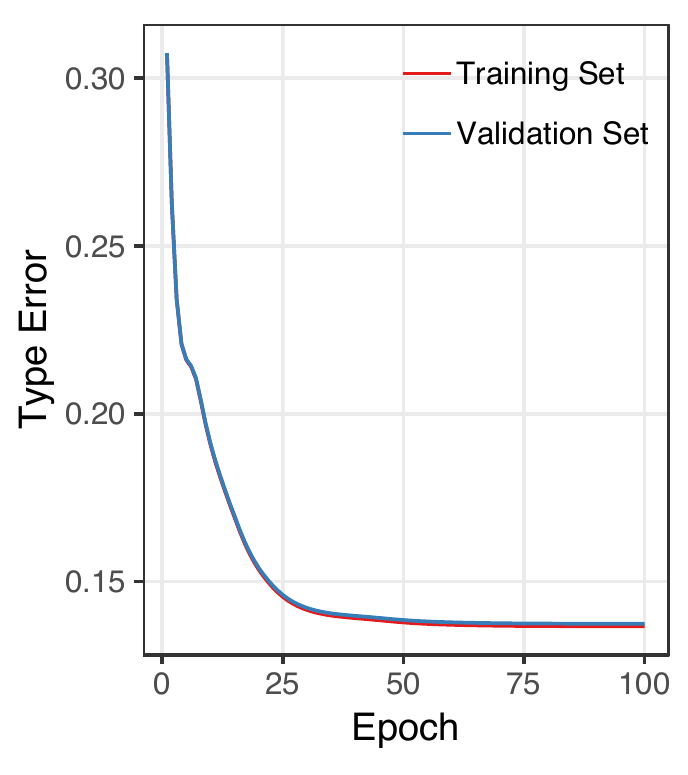}
\caption{Type error proportion.}
\label{fig:model-typeerr}
\end{subfigure}

\caption{Perplexity and type error proportion of the LSTM model measured
against the training and validation sets over epochs. They are averaged across
the 10 cross-validation sets.}
\label{fig:model-test}
\vspace{-1.0em}
\end{figure}

Figure~\ref{fig:model-test} illustrates the \textit{perplexity} and \textit{type
error} of the LSTM model measured on the training and validation sets. Recall
that the loss function of the model is a sum of the \textit{log perplexity} and
\textit{type error} (\S\ref{ss:phase2}).

\noindent\textbf{Perplexity.} Perplexity measures how well a trained model
predicts the next word that follows given words without perplexing. It is a
common metric for evaluating natural language models~\cite{
mikolov:acisca:2010, jozefowicz:arxiv:2016}.
A model with a lower perplexity performs better in predicting the next probable
fragment. Note from Figure~\ref{fig:model-pplx} that the perplexities for both
the training and validation sets decrease without a major difference as training
goes on.

\noindent\textbf{Type error.} Type error presents how well our model predicts
the correct type of a next fragment (recall \S\ref{ss:phase2}). A model with a
low type error is capable of predicting the fragments with the correct type in
its top predictions. Note from Figure~\ref{fig:model-typeerr} that the type
errors for both the training and validation sets  continuously  decrease and
become almost equal as the epoch increases.

The small differences of each perplexity and type error between the training set
and validation set demonstrate that our LSTM model is capable of learning the
compositional relations among fragments without overfitting or underfitting.

We further observed that epoch 70 is the optimal point at which both valid
perplexity and valid type errors start to plateau. We also noticed that the test
perplexity and test type errors at epoch 70 are 28.07 and 0.14, respectively.
Note from Figure~\ref{fig:model-test} that these values are close to those from
the validation set. It demonstrates that the model can accurately predict
fragments from the test set as well.
Thus, for the remaining  evaluations, we selected the model trained up to epoch
70, which took 6.6 hours on our machine.

\subsection{Effect of the \topk Parameter}
\label{ss:topk-test}

\input{tables/compare_table}

\sysname assembles model-suggested fragments when replacing an AST subtree of a
given JS code. In this process, \sysname randomly picks one fragment from the
\textit{Top-k} (\topk) suggestions for each insertion. Our key intuition is that
selecting fragments from the \textit{Top-k} rather than \textit{Top-1}
suggestion helps \sysname generate diverse code, which follows the pattern of JS
codes in our dataset but slightly differs from them. We evaluated the effect of
the \topk with seven different values varying from 1 to 64 to verify our
intuition.

\begin{figure}[!t]
\centering
  \includegraphics[width=\columnwidth]{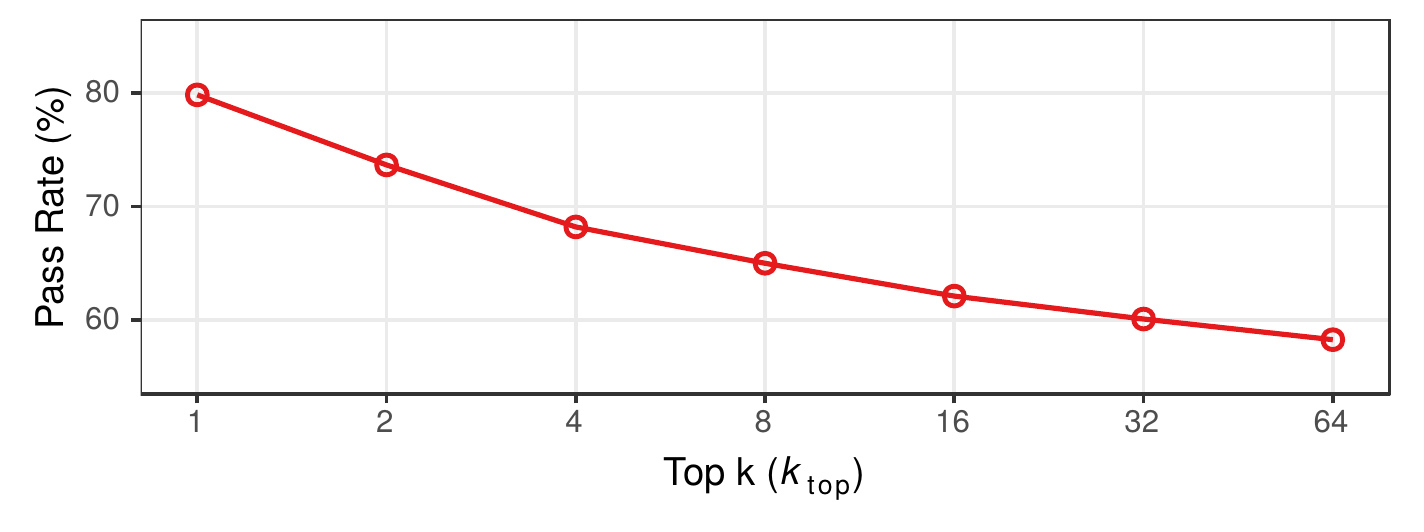}
\caption{The pass rate of generated JS tests over \topk.}
\label{fig:pr-test}
\vspace{-1.0em}
\end{figure}

We measured the pass rate of generated JS tests. A pass rate is a
measurement unit of demonstrating how many tests a target JS engine executes
without errors among generated test cases. To measure the pass rate, we first
generated 100,000 JS tests with each \topk value. We only considered five
runtime errors defined by the ECMAScript standard as errors~
\cite{ecma-script}. We then ran \sysname for 12 hours with each \topk value to
count the number of crashes found in \chakra 1.4.1.

\begin{figure}[!t]
\begin{subfigure}{0.49\columnwidth}
\centering
  \includegraphics[scale=0.54]{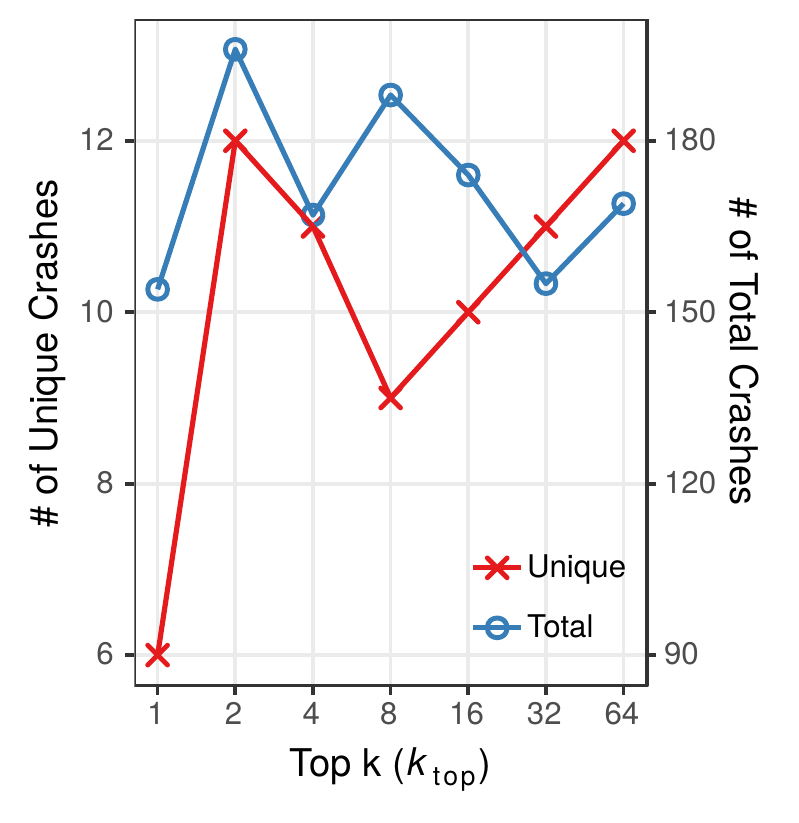}
\caption{Crashes on the release build.}
\label{fig:crash-release}
\end{subfigure}
\begin{subfigure}{0.49\columnwidth}
\centering
  \includegraphics[scale=0.54]{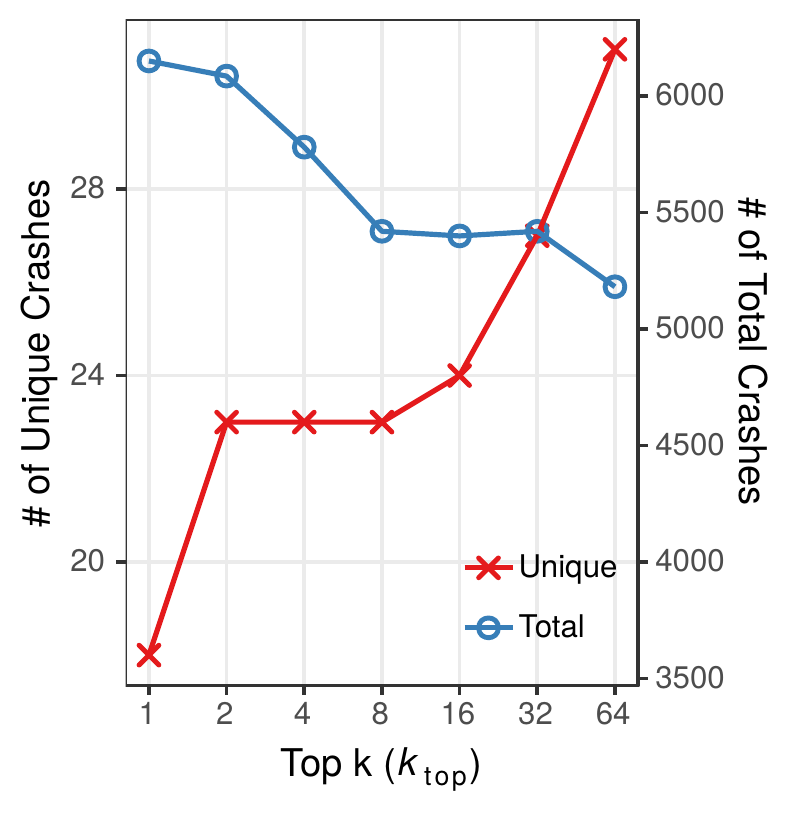}
\caption{Crashes on the debug build.}
\label{fig:crash-debug}
\end{subfigure}

\caption{The number of total and unique crashes found in \chakra 1.4.1 while
varying the \topk.
}
\label{fig:crash-test}
\vspace{-1.0em}
\end{figure}

Figures~\ref{fig:pr-test} and \ref{fig:crash-test} summarize our two
experimental results, respectively. As shown in Figure~\ref{fig:pr-test}, the
pass rate of \sysname decreases from 79.82\% to 58.26\% as the \topk increases.
This fact demonstrates that the suggestion from the model considerably affects
the generation of executable JS tests. It is also consistent with the
experimental results from Figure~\ref{fig:crash-debug}, in that \sysname finds
fewer total crashes when considering more fragment suggestions in generating JS
tests.
Note that Michael~\etal~\cite{patra:2016} demonstrated that their TreeFuzz
achieved a 14\% pass rate, which is significantly lower than that \sysname
achieved.

However, note from Figure~\ref{fig:crash-debug} that the number of unique
crashes increases, as \topk increases, unlike that of total crashes. This
observation supports our intuition that increasing the \topk helps \sysname
generate diverse JS tests that trigger undesired crashes in the JS engines.
Figure~\ref{fig:crash-test} also shows that \sysname found more crashes from the
debug build than the release build. Moreover, unlike the debug build, the
results for the release build did not show a consistent pattern. We believe
these results are mainly due to the nature of the debug build. It behaves more
conservatively with inserted assertion statements, thus producing crashes for
every unexpected behavior.

As Klees~\etal~\cite{klees:ccs:2018} stated, fuzzers should be evaluated
using the number of unique crashes, not that of crashing inputs. For both
release and debug builds of \chakra 1.4.1, \sysname found the largest number of
unique crashes when the \topk was 64. Therefore, we picked the \topk to be 64
for the remaining experiments.

\subsection{Comparison to State-of-the-art Fuzzers}
\label{ss:evalcmp}

To verify the ability to find bugs against open-source state-of-the-art
fuzzers, we compared \sysname with \alchemist~\cite{han:ndss:2019},
\jsfunfuzz~\cite{funfuzz}, and \ifuzzer~\cite{veggalam:esorics:2016}.
\jsfunfuzz and \ifuzzer have been used as a controlled group in the comparison
studies~\cite{han:ndss:2019, holler:usec:2012}. Furthermore, \alchemist, which
assembles its building blocks in a semantics-aware fashion, and \ifuzzer, which
employs an evolutionary approach with genetic programming, have in common with
\sysname in that they take in a corpus of JS tests.
Since \sysname, \alchemist, and \ifuzzer start from given seed JS files, we fed them
the same dataset collected from the repositories of Test262 and the four major JS
engines.
For fair comparison, we also configured \jsfunfuzz to be the version of January
31, 2017, on which we collected our dataset (recall \S\ref{ss:evalenv}).

We ran all four fuzzers on \chakra 1.4.1 and counted the number of found unique
crashes and known CVEs.
Since most fuzzers depend on random factors, which results in a high variance
of fuzzing results~\cite{klees:ccs:2018}, we conducted five trials; each trial
lasted for 6,336 CPU hours (72 hours $\times$ 88 cores).
We intentionally chose such a long timeout,  because fuzzers using evolutionary
algorithms, such as \ifuzzer, could improve their bug-finding ability as more tests
are generated.
Note that we expended a total of 31,680 CPU hours on the five trials of each
fuzzer.
Because \sysname took 6.6 hours to train its language model and used this model
for the five trials, we set the timeout of other fuzzers 1.3 hours (6.6 hours
$/$ 5 trials) longer than that of \sysname for fair comparison.

The \sysname, CA, \jsfunfuzz, and \ifuzzer columns of Table~\ref{tab:cmp-test}
summarize the statistical analysis of the comparison experimental results.
For the release build, \sysname found the largest number of CVEs, whereas
\jsfunfuzz still discovered more unique crashes than others.
For the debug build, \sysname outperformed all others in finding  both unique
crashes and CVEs.
We performed two-tailed Mann Whitney U tests and reported $p$-values between
\sysname and the other fuzzers in the table. We verified that all results are
statistically significant with $p$-values less than 0.05.

The last two rows of the table show the number of total and common bugs found in
the five trials from the release and debug builds, respectively. We counted common bugs when
\sysname found these bugs in every run of the five campaigns. When a bug was found
during at least one campaign, they are counted in the total bugs.
Note that \sysname found at least 2.14$\times$ more CVEs compared to others in
a total of the five trials. We believe that these results explain the
significance of \sysname in finding security bugs compared to the other
state-of-the-art fuzzers.

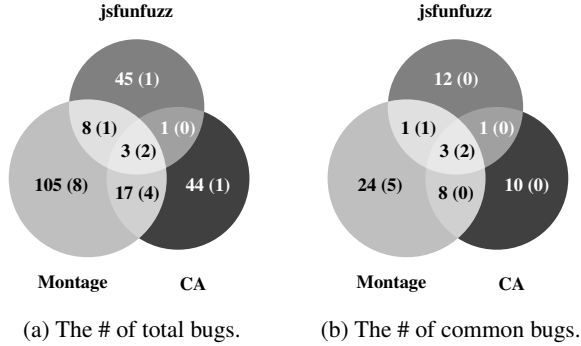
\begin{figure}[!t]
  \begin{subfigure}{0.49\columnwidth}
    \centering
    \input{figs/compare1}
    \caption{The \# of total bugs.}
    \label{f:cmp-total}
  \end{subfigure}
  \begin{subfigure}{0.49\columnwidth}
    \centering
    \input{figs/compare2}
    \caption{The \# of common bugs.}
    \label{f:cmp-common}
  \end{subfigure}
  \caption{The comparison of unique crashes (known CVEs) found by \sysname,
  \alchemist (CA), and \jsfunfuzz.}
  \label{f:cmp-bugs}
\vspace{-1em}
\end{figure}

We also compared the bugs discovered by each fuzzer.
Figure~\ref{f:cmp-bugs} depicts the Venn diagrams of unique bugs found in
\chakra 1.4.1.  These Venn diagrams present the total and common bugs that
each fuzzer found, corresponding to the last two rows of Table~\ref{tab:cmp-test}.
We excluded \ifuzzer from the figure because all found CVEs were also discovered
by \sysname.

Note from Figure~\ref{f:cmp-total} that \sysname identified 105 unique crashes
in total, including eight CVEs that were not found by \alchemist and \jsfunfuzz.
Furthermore, \sysname discovered all CVEs that were commonly found in the five
trials of \alchemist and \jsfunfuzz, as shown in Figure~\ref{f:cmp-common}.
However, \alchemist and \jsfunfuzz also identified a total of 45 and 46
unique bugs that were not found by \sysname, respectively.
These results demonstrate that \sysname plays a complementary role against the
state-of-the-art fuzzers in finding distinctive bugs.

\noindent\textbf{Performance over time.}
Figure~\ref{fig:bug-time}  shows the number of CVEs that \sysname found over time.
The number increases rapidly  in the first 1,144 CPU hours (13 hours $\times$ 88 cores) of the fuzzing
campaigns; however, \sysname finds additional bugs after running for
2,640 CPU hours (30 hours $\times$ 88 cores), thus becoming slow to find
new vulnerabilities.

\subsection{Effect of Language Models}

\begin{figure}[!t]
\centering
  \includegraphics[width=\columnwidth]{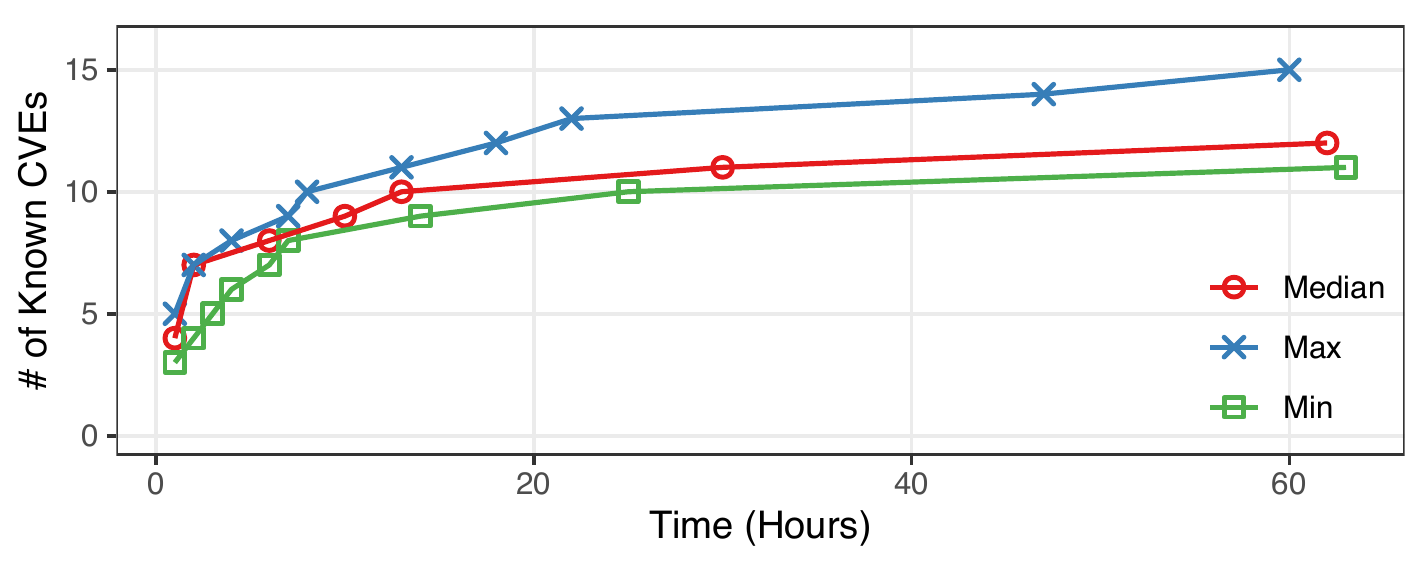}
\caption{The number of CVEs found by \sysname over time.}
\label{fig:bug-time}
\vspace{-1em}
\end{figure}

\label{ss:model-effectiveness}
\sysname generates JS tests by assembling language model-suggested fragments.
Especially, it takes advantage of the LSTM model to reflect the arbitrary length
of preceding fragments when predicting the next relevant fragments.
However, \sysname can leverage any other prevailing language models by its
design, and the language model it employs may substantially affect its fuzzing
performance.
Therefore, to analyze the efficacy of the LSTM model in finding bugs, we first
conducted a comparison study against two other approaches: (1) a random fragment
selection, and (2) Markov model-driven fragment selection.

The former approach is the baseline for \sysname where fragments are randomly
appended instead of querying a model. The latter approach uses a Markov model
that makes a prediction based on the occurrence history of the preceding two
fragments.
Specifically, we tailored the code from~\cite{markovify} to implement
the Markov chain.

Additionally, we compared our approach against a character/token-level RNN
language model-guided selection. It leverages an NNLM to learn the intrinsic
patterns from training instances, which is in common with ours.
Recently proposed approaches~\cite{godefroid:ase:2017, cummins:issta:2018,
liu:aaai:2019}, which resort to an NNLM to generate highly structured inputs,
adopted an approach more or less similar to this one.

Note that there is no publicly available character/token-level RNN model to
generate JS tests. Thus, we referenced the work of
Cummins~\etal~\cite{cummins:issta:2018} to implement this approach and trained
the model from scratch. To generate test cases from the trained model, we
referenced the work of Liu~\etal~\cite{liu:aaai:2019} because their approach is
based on the seed mutation like our approach.

The random, Markov, and ch-RNN columns of Table~\ref{tab:cmp-test} summarize the
number of crashes found by each approach. We conducted five fuzzing campaigns,
each of which lasted 72 hours; all the underlying experimental settings are
identical to those in \S\ref{ss:evalcmp}. Note that we conducted resolving
reference errors and fed the same dataset as \sysname when evaluating the
aforementioned three models.
\sysname outperformed the random selection and character/token-level RNN methods
in the terms of finding crashes and security bugs; thus, yielding $p$-values under 0.05,
which suggests the superiority of \sysname with statistical significance.

When comparing the metrics from release and debug build between \sysname and
the Markov chain approach, \sysname performed better. \sysname found more unique
bugs in total as well. However, the Mann Whitney U test deems the difference
insignificant.
Nevertheless, we emphasize that \sysname is capable of composing sophisticated
subtrees that the Markov chain easily fails to generate. For instance, \sysname
generated a JS test triggering CVE-2017-8729 by appending 54 fragments, which
the Markov chain failed to find. We provide more details of this case in
\S\ref{ss:case1}.

\begin{figure}[!t]
\centering
  \includegraphics[width=\columnwidth]{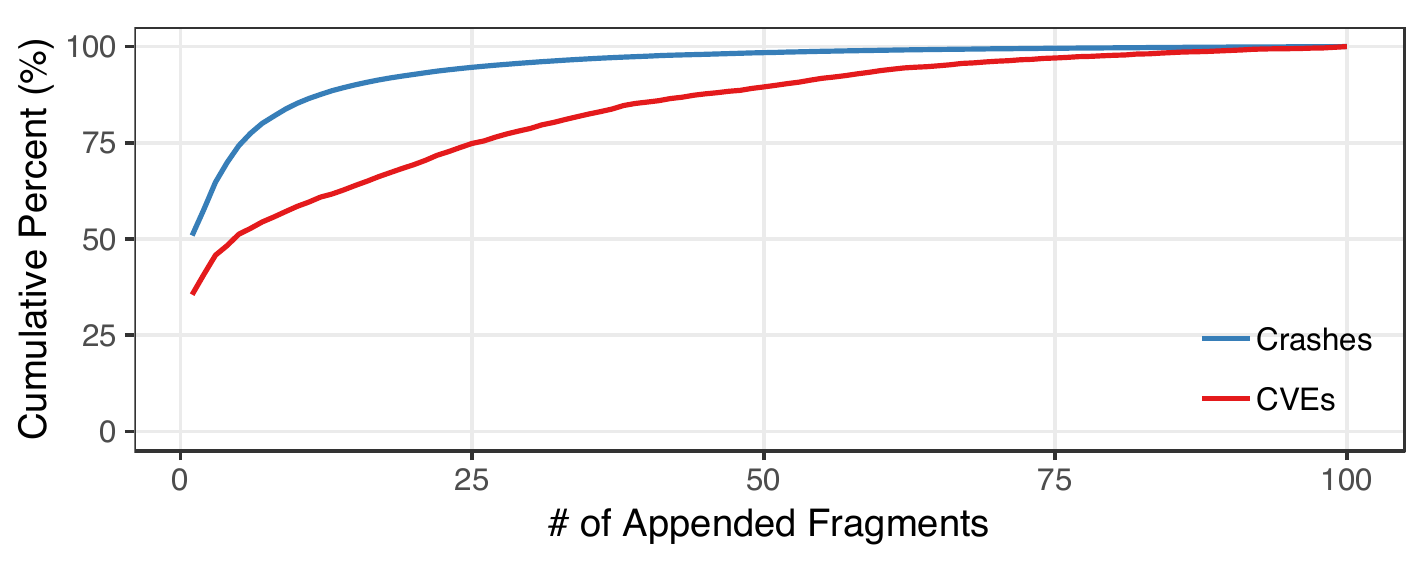}
  \caption{Empirical CDF of the number of appended fragments against JS tests
  causing crashes in \chakra.}
\label{fig:cdf-test}
\vspace{-1.0em}
\end{figure}

To evaluate the effectiveness of the LSTM model, we further analyzed the number
of fragments \sysname appended to generate JS tests that caused \chakra 1.4.1 to
crash in the experiment from \S\ref{ss:evalcmp}.

Figure~\ref{fig:cdf-test} shows the cumulative distribution function (CDF) of
the number of inserted fragments against 169,072 and 5,454 JS tests causing
crashes and known CVEs, respectively.
For 90\% of JS tests that caused the JS engine to crash, \sysname only assembled
fewer than 15 fragments; however, it appended up to 52 fragments to generate
90\% of JS tests that found the known CVEs.
This demonstrates that \sysname should append more fragments suggested by the
model to find security bugs rather than non-security bugs.
It also denotes that the random selection approach suffers from finding
security bugs.
Note that Table~\ref{tab:cmp-test} also accords with this result.

From the aforementioned studies, we conclude that the LSTM model trained on
fragments is necessary for finding bugs in the JS engines.

\noindent
\textbf{Resolving reference errors.}
We evaluated the importance of the reference error resolution step (recall
\S\ref{ss:error-resolution}) in finding JS engine bugs.
Specifically, we ran \sysname with the same settings as other approaches while
letting it skip the reference error resolution step but still leverage the same
LSTM model.
The last column of Table~\ref{tab:cmp-test} demonstrates that \sysname finds
fewer bugs if the resolving step is not applied, denoting that the error
resolving step improves the bug-finding capability of \sysname.
However, \sysname still found more bugs than the other state-of-the-art fuzzers
and the random approach even without the resolving step. Considering the random
approach also takes advantages of the error resolution step, the LSTM model
significantly contributes to finding JS engine bugs.

\begin{figure}[!t]
\centering
  \includegraphics[width=\columnwidth]{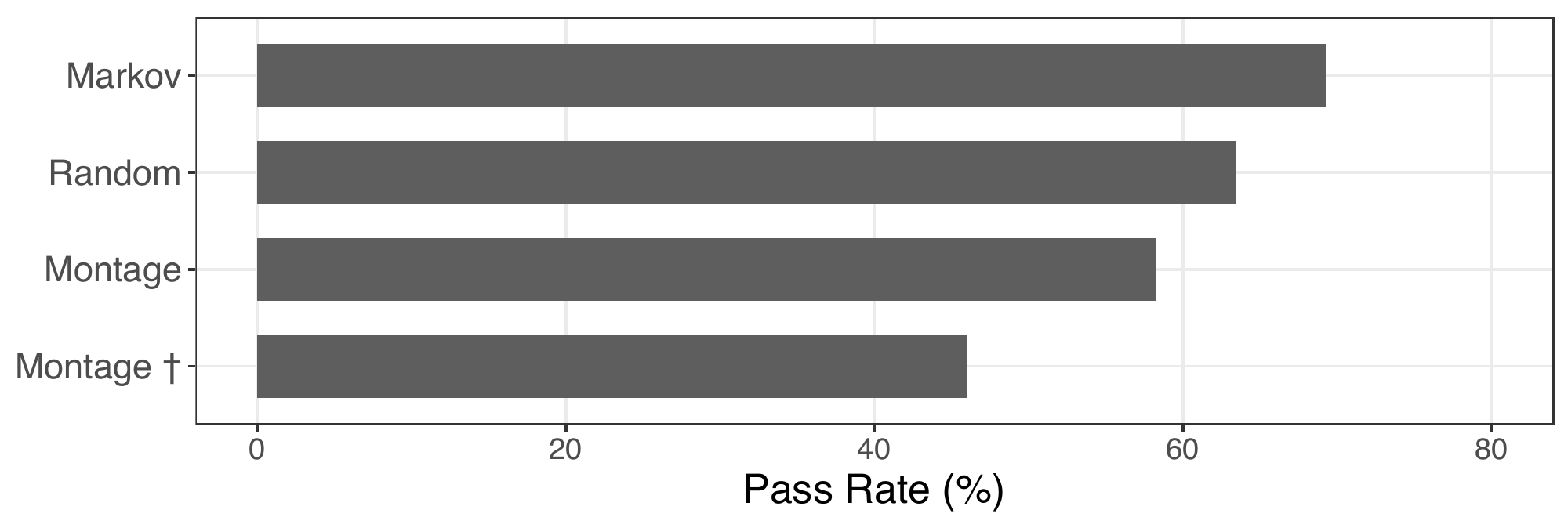}
  \caption{The pass rate measured against four different approaches: \sysname
  (\topk = 64) with and without resolving reference errors, random selection,
  and Markov model. \sysname without resolving reference errors is denoted by
  $\dagger$.}
  \label{fig:pr-cmp}
  \vspace{-1em}
\end{figure}

\noindent\textbf{Pass rate.} One of the key objectives of \sysname is to
generate a valid JS test so that it can trigger deep bugs in JS engines. Thus,
we further measured how the use of language models affects the pass rate of
generated codes.
A pass rate indicates whether generated test cases are indeed executed after
passing both syntax and semantic checking.

Figure~\ref{fig:pr-cmp} illustrates the pass rate of 100,000 JS tests generated
by the four different approaches: \sysname with and without resolving reference
errors, the random selection, and the Markov model.
We excluded the character/token-level RNN approach because only 0.58\% of the
generated tests were executed without errors. Such a low pass rate could be one
possible reason why this approach failed to find many bugs, as shown in
Table~\ref{tab:cmp-test}.
As Liu~\etal~\cite{liu:aaai:2019} also stated in their paper, we believe this
result is attributed to the lack of training instances and the unique
characteristics inherent in the regression test suite.

Note from the figure that resolving reference errors increases the pass rate by
12.2\%. As a result, this helped \sysname to find more bugs, as shown in
Table~\ref{tab:cmp-test}.
On the other hand, the pass rates of the random selection and Markov
model-guided approach were 5.2\% and 11\% greater than that of \sysname,
respectively.
We manually inspected the JS tests generated by the random selection and the
Markov model-guided approaches. We concluded that these differences stem from
appending a small number of fragments.
For instance, if a model always replaces one fragment, such as a string literal,
from the seed file, all generated JS tests will be executed without errors.

\subsection{Field Tests}\label{ss:evalfield}

We evaluated the capability of \sysname in finding real-world bugs. \textit{We
have run \sysname for 1.5 months on the latest production versions of the four
major engines:} \chakra, \jsc, \moz, and \veight.
For this evaluation, we collected datasets from the repository of each JS engine
at the version of February 3, 2019.
We additionally collected 165 PoCs that triggered known CVEs as our dataset.
Then, we trained the LSTM model for each JS engine.

\sysname has found \ourbug unique bugs from the four major JS engines so far.
Among the found bugs, 34 bugs were from \chakra. The remaining two and one bugs
were from \jsc and \veight, respectively.
We manually triaged each bug and reported all the found bugs to the
vendors. In total, \emph{\fixbug of the reported bugs have been patched so far}.

Especially, we reported three of the found bugs as security-related because they
caused memory corruptions of the target JS engines. The three security bugs were
discovered in \chakra 1.11.7, \chakra 1.12.0 (beta), and \jsc 2.23.3,
respectively.
Note that all of them got CVE IDs: CVE-2019-0860, CVE-2019-0923, and
CVE-2019-8594.
Particularly, we were rewarded for the bugs found in \chakra with a bounty of
5,000 USD.

Our results demonstrate that \sysname is capable of finding \ourbug real-world
JS engine bugs, including three security bugs. We further describe one of the
real-world security bugs that \sysname found in \S\ref{ss:case3}.

\subsection{Case Study}\label{ss:casestudy}

To show how \sysname leverages the existing structure of the regression test, we
introduce three bugs that \sysname found.
We show two bugs that \sysname found in \chakra 1.4.1 from the experiment in
\S\ref{ss:evalcmp}. We then describe one of the real-world security bugs
found in the latest version of \chakra, which is already patched. Note that we
minimized all test cases for ease of explanation.

\subsubsection{CVE-2017-8729} \label{ss:case1}

\begin{figure}[ht]
  \begin{lstlisting}[escapechar=$]
  (function () {
      for (var v0 in [{ $\label{line:bodystart}$
          v1 = class {},
          v2 = 2010
      }.v2 = 20]) { $\label{line:patternend}$
          for([] in {
              value: function() {},
              writable: $\color{superdarkblue}{false}$
          }){}
      } $\label{line:bodyend}$
  })();
  \end{lstlisting}
  \caption{A test code that triggers CVE-2017-8729 on \chakra
  1.4.1.}
  \label{lst:case1}
\end{figure}

Figure~\ref{lst:case1} shows the minimized version of a generated test that
triggers CVE-2017-8729 on \chakra 1.4.1.
From its seed file, \sysname removed the body of \texttt{FunctionExpression}
and composed a new subtree corresponding to
Lines~\ref{line:bodystart}-\ref{line:bodyend}.
Particularly, \sysname appended 54 fragments to generate the new test.

\chakra is supposed to reject the generated test before its execution because it
has a syntax error in the \texttt{ObjectPattern} corresponding to
Lines~\ref{line:bodystart}-\ref{line:patternend}.
However, assuming the \texttt{ObjectPattern} to be an \texttt{ObjectExpression},
\chakra successfully parses the test and incorrectly infers the type of the
property \texttt{v2} to be a setter. Thus, the engine crashes with a
segmentation fault when it calls the setter in Line~\ref{line:patternend}.
Interestingly, the latest version of \chakra still executes this syntax-broken
JS test without errors but does not crash.

The original regression test checked the functionalities regarding a complicated
\texttt{ObjectPattern}. Similarly, the generated test triggered a type confusion
vulnerability while parsing the new \texttt{ObjectPattern}.
Therefore, we believe that this case captures the design goal of \sysname, which
leverages an existing regression test and puts it in a different execution
context to find a potential bug.

\subsubsection{CVE-2017-8656}

\begin{figure}[ht]
  \begin{lstlisting}[escapechar=$]
  var v1 = {
      'a': function () {}
  }
  var v2 = 'a';
  (function () {
      try {
      } catch ([v0 = (v1[v2].__proto__(1, 'b'))]) { $\label{line:catch}$
          var v0 = 4; $\label{line:decl}$
      }
      v0++; $\label{line:crash}$
  })();
  \end{lstlisting}
  \caption{A test code that triggers CVE-2017-8656 on \chakra
  1.4.1.}
  \label{lst:case2}
\end{figure}

Figure~\ref{lst:case2} shows a test case generated by \sysname that triggers
CVE-2017-8656 on \chakra 1.4.1. Its seed file had a different
\texttt{AssignmentExpression} as the parameter of a \texttt{CatchClause} in
Line~\ref{line:catch}.
From the seed AST, \sysname removed a subtree corresponding to the
\texttt{AssignmentExpression} in Line~\ref{line:catch} and mutated it by
appending eight fragments that the LSTM model suggested.

In the generated code, the variable \texttt{v0} is first declared as the
parameter of the \texttt{CatchClause} (Line~\ref{line:catch}) and then
redeclared in its body (Line~\ref{line:decl}). At this point, the \chakra
bytecode generator becomes confused with the scope of these two variables and
incorrectly selects which one to initialize. Consequently, the variable
\texttt{v0} in Line~\ref{line:decl} remains uninitialized. As the JS engine
accesses the uninitialized symbol in Line~\ref{line:crash}, it crashes with a
segmentation fault.

We note that the seed JS test aimed to check possible scope confusions, and the
generated code also elicits a new vulnerability while testing a  functionality
similar to the one its seed JS test checks. Hence, this test case fits the design
objective of \sysname.

\subsubsection{CVE-2019-0860} \label{ss:case3}

\begin{figure}[ht]
  \begin{lstlisting}[escapechar=$]
  function f0(f, p = {}) {
      f.__proto__ = p; $\label{line:assignprop}$
      f.arguments = 44; $\label{line:assignarg}$
      f.arguments === 44; $\label{line:binop}$
  }

  let v1 = new $\color{superdarkblue}{Proxy}$({}, {});
  for (let v0 = 0; v0 < 1000; ++v0) { $\label{line:loop-start}$
      f0(function () {'use strict';}, v1);
      f0(class C {}, v1);
  } $\label{line:loop-end}$
  \end{lstlisting}
  \caption{A test code that triggers CVE-2019-0860 on \chakra
  1.12.0 (beta).}
  \label{lst:case3}
\vspace{-1.0em}
\end{figure}

Figure~\ref{lst:case3} describes a JS test triggering CVE-2019-0860 on \chakra
1.12.0 (beta), which we reported to the vendor.
Its seed file had a \texttt{CallExpression} instead of the statements in
Lines~\ref{line:assignarg}-\ref{line:binop}. From the seed JS test, \sysname
removed a subtree corresponding to the \texttt{BlockStatement} of the function
\texttt{f0} and appended 19 fragments to compose a new block of statements
(Lines~\ref{line:assignprop}-\ref{line:binop}). Notably, \sysname revived the
\texttt{AssignmentExpression} statement in Line~\ref{line:assignprop}, which is
a required precondition to execute the two subsequent statements and trigger the
security bug.

The seed regression test was designed to test whether JS engines correctly
handle referencing the \texttt{arguments} property of a function in the strict
mode. For usual cases, JS engines do not allow such referencing; however, to
place the execution context in  an exceptional case, the seed JS test enables
the access by adding a \texttt{Proxy} object to the prototype chain of the function
\texttt{f} (Line~\ref{line:assignprop}). As a result, this new test is able to access and
modify the property value without raising a type error
(Line~\ref{line:assignarg}).

While performing the JIT optimization process initiated by the \texttt{for} loop
in Lines~\ref{line:loop-start}-\ref{line:loop-end}, \chakra misses a
postprocessing step of the property in Line~\ref{line:assignarg}, thus enabling
to write an arbitrary value on the memory. Consequently, the engine crashes
with a segmentation fault as this property is accessed in Line~\ref{line:binop}.

Note that the generated test case triggers a new vulnerability while vetting the
same functionality that its seed tests. Moreover, the \texttt{GlobOpt.cpp} file,
which is the most frequently patched file to fix CVEs assigned to \chakra, was
patched to fix this vulnerability. Therefore, this JS test demonstrates that
\sysname successfully discovers bugs that it aims to find.


%% file: tables/compare_table.tex
\newcolumntype{Y}{>{\centering\arraybackslash}X}

\begin{table*}
  \begin{threeparttable}
    \footnotesize
    \centering
    \vspace{-1.0em}
    \caption{The number of bugs found with four fuzzers and four different
    approaches: \sysname, \alchemist (CA), \jsfunfuzz, and \ifuzzer; random
    selection, Markov chain, char/token-level RNN, and \sysname (\topk = 64)
    without resolving reference errors. We marked results in bold when the
    difference between \sysname and the other approach is statistically significant.}
    \label{tab:cmp-test}
    \begin{tabularx}{\textwidth}{llY|YYY|YYY|Y}
      \toprule
        \multirow{2}{*}{\textbf{\vspace{-0.6em}Build}} &
        \multirow{2}{*}{\textbf{\vspace{-0.6em}Metric}} &
        \multicolumn{8}{c}{\textbf{\# of Unique Crashes (Known CVEs)}} \\
      \cmidrule{3-10}
        & & \textbf{\sysname} & \textbf{CA} & \textbf{\jsfunfuzz} & \textbf{\ifuzzer} & \textbf{random} & \textbf{Markov} & \textbf{ch-RNN} & \textbf{\sysname~\tnote{$\dagger$}} \\
      \midrule
        \multirow{7}{*}{Release} & Median & 23 (7) & 15 (4) & 27 (3) & 4 (1) & 12 (3) & 19 (6) & 1 (0) & 12 (4) \\
        & Max & 26 (8) & 15 (4) & 31 (4) & 4 (2) & 15 (4) & 22 (7) & 1 (1) & 13 (5) \\
        & Min & 20 (6) & 14 (3) & 25 (3) & 0 (0) & 10 (3) & 16 (5) & 0 (0) & 11 (4) \\
        & \multirow{2}{*}{Stdev} & 2.30 & 0.55 & 2.19 & 1.79 & 2.07 & 2.39 & 0.45 & 0.84 \\
        & & (0.84) & (0.55) & (0.45) & (0.71) & (0.45) & (0.84) & (0.55) & (0.45) \\
        & \multirow{2}{*}{$p$-value} & \multirow{2}{*}{N/A} & \textbf{0.012} & \textbf{0.029} & \textbf{0.012} & \textbf{0.012} & \textbf{0.037} & \textbf{0.012} & \textbf{0.012} \\
        & & & \textbf{(0.012)} & \textbf{(0.012)} & \textbf{(0.012)} & \textbf{(0.012)} & (0.144) & \textbf{(0.012)} & \textbf{(0.012)} \\
      \midrule
        \multirow{7}{*}{Debug} & Median & 49 (12) & 26 (6) & 27 (4) & 6 (1) & 31 (7) & 44 (11) & 3 (0) & 41 (9) \\
        & Max & 52 (15) & 30 (6) & 29 (5) & 8 (3) & 34 (7) & 50 (12) & 4 (1) & 43 (10) \\
        & Min & 45 (11) & 24 (4) & 24 (4) & 2 (0) & 27 (6) & 42 (8) & 1 (0) & 38 (8)\\
        & \multirow{2}{*}{Stdev} & 2.70 & 2.61 & 2.12 & 2.41 & 2.88 & 3.27 & 1.10 & 1.82 \\
        & & (1.64) & (0.89) & (0.45) & (1.10) & (0.45) & (1.67) & (0.5) & (0.84) \\
        & \multirow{2}{*}{$p$-value} & \multirow{2}{*}{N/A} & \textbf{0.012} & \textbf{0.012} & \textbf{0.012} & \textbf{0.012} & 0.144 & \textbf{0.012} & \textbf{0.012} \\
        & & & \textbf{(0.012)} & \textbf{(0.012)} & \textbf{(0.012)} & \textbf{(0.012)} & (0.298) & \textbf{(0.012)} & \textbf{(0.012)} \\
      \midrule
        \multirow{2}{*}{Both} & Total & 133 (15) & 65 (7) & 57 (4) & 22 (3) & 72 (9) & 109 (14) & 10 (2) & 74 (10) \\
        & Common & 36 (8) & 22 (2) & 17 (3) & 1 (0) & 29 (6) & 37 (8) & 1 (0) & 37 (7) \\
      \bottomrule
    \end{tabularx}
    \begin{tablenotes}
       \item[$\dagger$] \sysname without resolving reference errors.
    \end{tablenotes}
  \end{threeparttable}
\end{table*}

%% file: figs/compare1.tex
\begin{tikzpicture}
  \def\ours{(0,0) circle (1.05)}
  \def\funfuzz{(0.65,0.95) circle (0.9)}
  \def\alchemist{(1.2,0) circle (0.95)}
  \begin{scope}[blend group = screen]
    \fill[lightgray] \ours;
    \fill[gray] \funfuzz;
    \fill[darkgray] \alchemist;
  \end{scope}

  \node [font=\scriptsize] at (-0.35,-0.1) {\textbf{105 (8)}} ;
  \node [font=\scriptsize, text=white] at (1.6, -0.1) {\textbf{44 (1)}};
  \node [font=\scriptsize, text=white] at (0.66, 1.3) {\textbf{45 (1)}};
  \node [font=\scriptsize] at (0.16, 0.65) {\textbf{8 (1)}};
  \node [font=\scriptsize] at (0.66,-0.21) {\textbf{17 (4)}};
  \node [font=\scriptsize, text=white] at (1.2, 0.65) {\textbf{1 (0)}};
  \node [font=\scriptsize] at (0.68, 0.35) {\textbf{3 (2)}};
  \node [font=\scriptsize] at (0.65, 2.2) {\textbf{\jsfunfuzz}};
  \node [font=\scriptsize] at (-0.2, -1.4) {\textbf{\sysname}};
  \node [font=\scriptsize] at (1.4, -1.4) {\textbf{CA}};
\end{tikzpicture}

%% file: figs/compare2.tex
\begin{tikzpicture}
  \def\ours{(0,0) circle (1.05)}
  \def\funfuzz{(0.65,0.95) circle (0.9)}
  \def\alchemist{(1.2,0) circle (0.95)}
  \begin{scope}[blend group = screen]
    \fill[lightgray] \ours;
    \fill[gray] \funfuzz;
    \fill[darkgray] \alchemist;
  \end{scope}

  \node [font=\scriptsize] at (-0.35,-0.1) {\textbf{24 (5)}} ;
  \node [font=\scriptsize, text=white] at (1.6, -0.1) {\textbf{10 (0)}};
  \node [font=\scriptsize, text=white] at (0.66, 1.3) {\textbf{12 (0)}};
  \node [font=\scriptsize] at (0.16, 0.65) {\textbf{1 (1)}};
  \node [font=\scriptsize] at (0.66,-0.21) {\textbf{8 (0)}};
  \node [font=\scriptsize, text=white] at (1.2, 0.65) {\textbf{1 (0)}};
  \node [font=\scriptsize] at (0.68, 0.35) {\textbf{3 (2)}};
  \node [font=\scriptsize] at (0.65, 2.2) {\textbf{\jsfunfuzz}};
  \node [font=\scriptsize] at (-0.2, -1.4) {\textbf{\sysname}};
  \node [font=\scriptsize] at (1.4, -1.4) {\textbf{CA}};
\end{tikzpicture}

%% file: related.tex
\section{Related Work}

\noindent
\textbf{Fuzzing.}
There have been a vast volume of research on generation-based fuzzing.
Highly-structured file fuzzing~\cite{godefroid:pldi:2008,pham:ase:2016},
protocol fuzzing~\cite{ruiter:usec:2015, somorovsky:ccs:2016}, kernel
fuzzing~\cite{syzkaller,trinity,han:ccs:2017}, and interpreter fuzzing~\cite{funfuzz,
wang:oakland:2017, dewey:ase:2014, patra:2016, aschermann:ndss:2017,
blaz:usec:2019} are representative research examples.

IMF infers the model of sequential kernel API calls to fuzz macOS kernels~\cite{han:ccs:2017}.
Dewey~\etal~\cite{dewey:ase:2014} generated code with specified combinations of
syntactic features and semantic behaviors by constraint logic programming.

Godefroid~\etal~\cite{godefroid:ase:2017} trained a language model from a large
number of PDF files and let the model learn the relations between objects
constituting the PDF files. Their approach of using a language model in
generating tests is similar to ours \emph{per se}, but their approach is not
directly applicable to generating JS tests, which demands modeling
complicated control and data dependencies.

Cummins~\etal~\cite{cummins:issta:2018} also proposed a similar approach. They
trained an LSTM language model from a large corpus of OpenCL code. Unlike
\sysname, they trained the model at a character/token-level, which does not
consider the compositional relations among the AST subtrees.

TreeFuzz~\cite{patra:2016} is another model-based fuzzer. Its model is built
on the frequencies of co-occurring nodes and edges from given AST examples.
Their modeling of generating tests is not directly applicable to the prevalent
state-of-the-art language models, tailored to train word sequences, not node
and edge relations in ASTs.

Aschermann~\etal~\cite{aschermann:ndss:2017} and Blazytko~\etal~\cite{blaz:usec:2019}
recently proposed \nautilus and \grimoire, respectively. Both fuzzers test programs
that take highly structured inputs by leveraging code coverage. Based on a given grammar,
\nautilus generates a new JS test and checks whether it hits new code coverage for
further mutation chances.
Contrary to \nautilus, \grimoire requires no user-provided components, such
as grammar specification and language models, but synthesizes inputs that
trigger new code coverage. As they stated, \grimoire has difficulties in
generating inputs with complex structures, requiring semantic information.

Previous studies of mutational fuzzing~\cite{aflfuzz, woo:ccs:2013, rebert:usec:2014,
cha:oakland:2015,holler:usec:2012,guo:2013, veggalam:esorics:2016, liu:aaai:2019}
focus on altering given seeds to leverage functionalities that the seeds already test.

LangFuzz~\cite{holler:usec:2012} is a mutational fuzzing tool that substitutes
a non-terminal node in a given AST with code fragments. It iteratively
replaces non-terminal nodes in the step of inserting fragments. However,
LangFuzz does not consider any context regarding picking a promising candidate
to cause a target JS engine crash. On the other hand, \sysname is capable of learning
implicit relations between fragments that may be inherent in given examples.

Liu~\etal~\cite{liu:aaai:2019} proposed a mutation-based approach to fuzz the
target program. Given a large corpus of C code, they trained a
sequence-to-sequence model to capture the inherent pattern of input at
character-level. Then, they leveraged the trained model to mutate the seed.
Their approach suffers from the limitation that the model generates many malformed
tests, such as unbalanced parenthesis.

\noindent
\textbf{Language model for code.}
Hindle~\etal~\cite{hindle:icse:2012} measured the naturalness of software
by computing the cross-entropy values over lexical code tokens in large
JAVA and C applications. They also first demonstrated even count-based
n-gram language models are applicable to code completion. To make more
accurate suggestions for code completion, SLAMC~\cite{nguyen:fse:2013}
incorporated semantic information, including type, scope, and role for each
lexical token. SLANG~\cite{raychev:pldi:2014} lets a model learn API call sequences from
Android applications. It then uses such a model to improve the precision of
code completion. GraLan learns the relations between API calls from the
graph of API call sequences, and ASTLan uses GraLan to fill holes in the
AST to complete the code~\cite{nguyen:icse:2015}.

Maddison~\etal~\cite{maddison:icml:2014} studied the generative models of
natural source code based on PCFGs and source code-specific structures.
Bielik~\etal~\cite{bielik:icml:2016} suggested a new generative probabilistic
model of code called a probabilistic higher order grammar, which generalizes
PCFGs and parameterizes the production rules on a context.

The objective of using a language model in all of the above works is to make
better suggestions for code completion. However, \sysname focuses on generating
JS tests that should be accepted by a target JS engine.

%% file: conclusion.tex
\section{Conclusion}

We present \sysname, the first fuzzing tool that leverages an NNLM in generating
JS tests. We propose a novel algorithm of modeling the hierarchical structures
of a JS test case and the relationships among such structures into a sequence
of fragments. The encoding of an AST into a fragment sequence enables \sysname to
learn the relationships among the fragments by using an LSTM model. \sysname
found \ourbug real-world bugs in the latest JS engines, which demonstrates its
effectiveness in finding JS engine bugs.